\documentclass[journal]{IEEEtran}
\pdfoutput=1
\usepackage{lipsum}
\DeclareUnicodeCharacter{2212}{-}
\usepackage{multirow}
\usepackage{verbatimbox}
\usepackage{caption}
\DeclareCaptionLabelSeparator{none}{ }
\usepackage{placeins}
\usepackage{tabularx}
\usepackage{afterpage,lipsum}
\usepackage{float}
\usepackage{dblfloatfix}
\usepackage{algorithm}
\usepackage[noend]{algpseudocode}
\makeatletter
\usepackage{color,soul}

\usepackage{graphicx}

\def\BState{\State\hskip-\ALG@thistlm}
\makeatother
\newcommand*{\rttensortwo}[1]{\bar{\bar{#1}}}
\usepackage{stackengine}
\newcommand\xrowht[2][0]{\addstackgap[.5\dimexpr#2\relax]{\vphantom{#1}}}
\usepackage{xcolor}
\usepackage{colortbl}
\usepackage{breqn}
\usepackage[justification=centering]{caption}
\usepackage{amssymb}
\usepackage{subcaption}
\usepackage{bm}

\DeclareMathOperator*{\minimize}{{minimize}}

\usepackage{amsmath}

\usepackage{mathtools}
\usepackage{amsfonts}
\usepackage{gensymb}
\usepackage{hyperref}
\ifCLASSINFOpdf
\else
\fi

\begin{document}

\title{A Combined Machine-Learning / Optimization-Based Approach for Inverse Design of Nonuniform Bianisotropic Metasurfaces}

\author{Parinaz Naseri, Stewart Pearson, Zhengzheng Wang,
        and~Sean~V. Hum \\
         Submitted to IEEE Transactions on Antennas \& Propagation, \\ Machine Learning in Antenna Design, Modeling, and Measurements Special Issue
        
\thanks{P. Naseri, S. Pearson, Z. Wang, and S. V. Hum are with the Edward S. Rogers Sr. Department of Electrical and Computer Engineering, 10 King's College Road, toronto, Ontario, Canada, M5S3G4, email: parinaz.naseri@utoronto.ca }}

\markboth{IEEE TRANSACTIONS ON ANTENNAS AND PROPAGATION, VOL. , NO. ,  2021}%
{Shell \MakeLowercase{\textit{et al}}: Bare Demo of IEEEtran.cls for IEEE Journals}

\maketitle

\begin{abstract}
Electromagnetic metasurface design based on far-field constraints without the complete knowledge of the fields on both sides of the metasurface is typically a time consuming and iterative process, which relies heavily on heuristics and \textit{ad hoc} methods. This paper proposes an end-to-end systematic and efficient approach where the designer inputs high-level far-field constraints such as nulls, sidelobe levels, and main beam level(s); and a 3-layer nonuniform passive, lossless, omega-type bianisotropic electromagnetic metasurface design to satisfy them is returned. The surface parameters to realize the far-field criteria are found using the alternating direction method of multipliers on a homogenized model derived from the method of moments. This model incorporates edge effects of the finite surface and mutual coupling in the inhomogenous impedance sheet. Optimization through the physical unit cell space integrated with machine learning-based surrogate models is used to realize the desired surface parameters from physical meta-atom (or unit cell) designs. Two passive lossless examples with different feeding systems and far-field constraints are shown to demonstrate the effectiveness of this method.
\end{abstract}


\IEEEpeerreviewmaketitle

\section{Introduction}

\IEEEPARstart{E}{lectromagnetic} metasurfaces (EMMSs) are thin uniform or nonuniform 2D arrangements of sub-wavelength unit cells that are composed of patterned metallic scatterers and/or dielectric substrates. These special surfaces provide the ability to manipulate electromagnetic waves in extraordinary ways. Examples include spectrum filtering, wave manipulation, and polarization conversion \cite{OQT}. The \textit{inverse} design of an EMMS generally involves two steps: mapping the constraints on the fields of one side of the surface while knowing the fields on the other side of the EMMS to macroscopic properties; and then, achieving these properties microscopically using physical unit cells. The EMMS macroscopic electromagnetic properties can be described in terms of the scattering parameters, surface susceptibilities \cite{Achouri}, surface impedance/admittance \cite{Epstein}--\cite{FF0}, or {equivalent permittivity and permeability tensors} \cite{Ol1}--\cite{NaseriTA}. It is shown that bianisotropy, which can be modeled by a magneto-electric coupling term ${K}_{em}$, is also needed for useful wave transformations \cite{Epstein}--\cite{Chen}. Since these macroscopic parameters describe the electromagnetic properties of the same EMMS, they are equivalent and interchangeable. Here, we use the surface electric impedance ($\rttensortwo{Z}_{se}$), surface magnetic admittance ($\rttensortwo{Y}_{sm}$), and electro-magnetic coupling coefficients ($\rttensortwo{K}_{em}$) to describe the surface properties of the EMMS.

\begin{figure}[h]
  \centering 
    \includegraphics[width=3.6in]{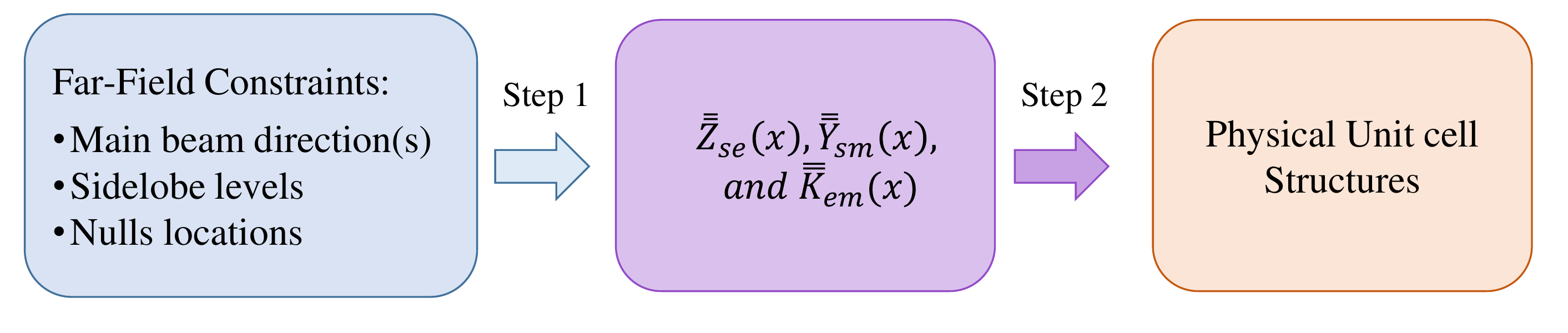}
    \caption{The steps to inverse design a metasurface based on the constraints on the desired radiated far-field, based on the the surface impedance approach.. }
    \label{fig1}
\end{figure}

The macroscopic and microscopic steps of the \textit{inverse problem} are shown with steps 1 and 2 in Figure \ref{fig1}, respectively. In practical and general problems where the description of the field on one side is incomplete or high level constraints on the desired radiation pattern are given, the surface properties need to be obtained without the knowledge of the tangential electric and magnetic fields on one side of the surface. Hence, the first step of the \textit{inverse problem} can no longer be solved analytically and requires optimization to search for the surface parameters \cite{FF01}--\cite{FF5}. Moreover, the design of the right unit cell structure based on the macroscopic surface parameters is a one-to-many mapping where multiple solutions may exist in a high-dimensional design space. This makes the search for the right design difficult and has resulted in this step being implemented by mostly \textit{ad hoc} or empirical approaches in the literature. So far, this approach relies on time-consuming and resource-demanding cycles of optimization and full-wave simulations, that has been the bottleneck of the EMMS design process. 

Interesting power-conserving wave transformations are possible with nonuniform surfaces where the surface parameters are a function of position on the EMMS surface. These parameters translate to a nonuniform array of unit cells that are composed of at least three layers of asymmetric patterned metallic scatterers \cite{Epstein} in the physical solution domain. It is worth noting that the surface properties of each unit cell are characterized by simulating them in periodic boundary conditions. Therefore, when placed in a nonuniform array, the unit cells' effective properties can vary from their designed values. This change is attributed to mutual coupling between each unit cell with its neighbors. Neglecting to account for this coupling results in errors in the wave transformation. The designer is then required to simulate and tune each of the many variables such as scatterers' dimensions to obtain the desired response. This \textit{ad hoc} approach requires many cycles of optimization through expensive full-wave simulations and is very time- and resource-demanding. For transverse electric (TE) incident waves, the use of vertical baffles to implement perfect conductor walls between neighbor unit cells has been proposed to suppress the mutual coupling \cite{Paul}, but it complicates the fabrication and only works for TE excitation.

The second step involves optimizing the unit cell structure based on the required macroscopic properties. Designers mostly rely on empirical approaches including many cycles of optimization and simulations to obtain the right choices for the unit cells' scatterer shapes, dimensions, substrate thicknesses and permittivity(ies). Searching for the optimized structure is a challenging problem since the solution space includes a wide variety of scatterer shapes and corresponding feature dimensions. This solution space might include many local minima, which make the optimization over this space difficult. Moreover, to find the right choice, each candidate must be simulated and evaluated against the required properties. Inspired by the revolution that data-driven machine-learning methods have made in material informatics such as discovery of new quantum materials, pharmaceuticals, and other compounds \cite{VAE}, deep  machine learning \cite{arrebola}--\cite{b7} and statistical learning \cite{Ol2} methods can help to build surrogate models that can provide fast predictions of the properties of each unit cell. Moreover, several machine-learning techniques have been proposed to tackle the challenges of the second step \cite{b10}--\cite{Tandum}. Some of these proposed methods deal with the inverse design of a uniform EMMS where the impact of mutual coupling is less of an issue \cite{b10}--\cite{b20}, while the rest optimize over a simple solution space that is composed of only one scatterer shape \cite{b8}--\cite{Tandum}. It is worth noting that dielectric optical EMMSs can be designed using global optimization methods due to the analytical relation between the scatterers' properties and the EMMS's scattering parameters  \cite{GLOnet}. However, due to the lack of such relations in EMMSs composed of metallic scatterers, the inverse design of a heterogeneous bianisotropic EMMS is more challenging. Nonetheless, solving this problem more efficiently has led to the proposal of systematic approaches that provide both the optimized surface properties and the actual physical unit cells to synthesize a stack of two EMMSs that generates two pencil beams at two frequency bands \cite{Grbic2021}.

\begin{figure}[!ht]
	\centering
  	\includegraphics[width=3.in]{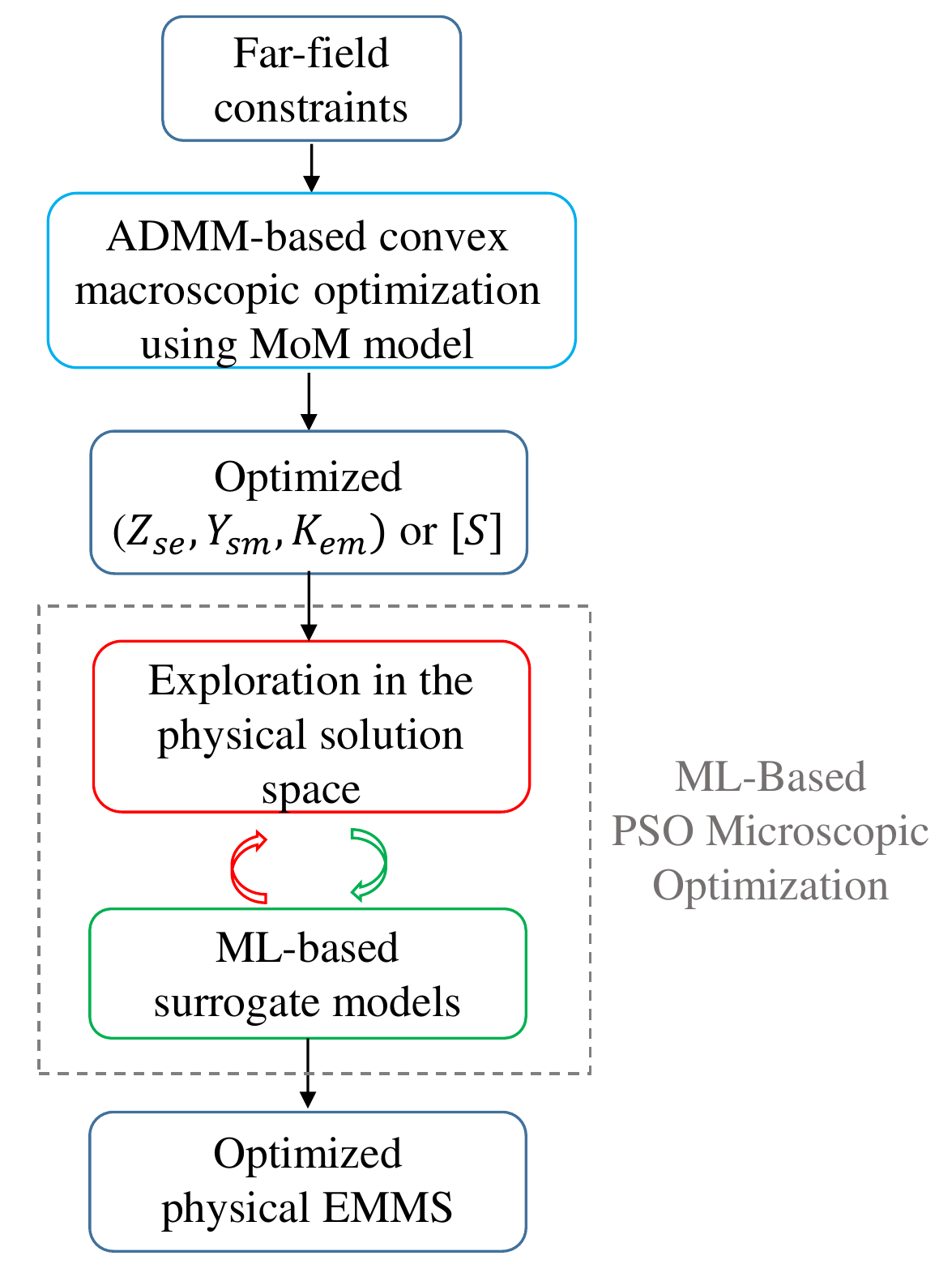}
  	\caption{Proposed systematic approach for inverse design of a nonuniform bianisotropic metasurface based on the constraints on the radiated far-field pattern.}
  	\label{fig:flowchart}
\end{figure}

Here, to expand the applications of the EMMSs, arbitrary far-field constraints are given as the inputs and the macroscopic and microscopic steps of the inverse problem of the EMMSs are combined together into a unified design approach that helps to implement a nonuniform bianisotropic metasurface to satisfy them. Figure \ref{fig:flowchart} shows the flowchart of the proposed systematic approach. This approach allows a designer to place practical constraints on the radiated far-zone fields emanating from the physical implementation of the EMMS in an efficient and systematic way. In the macroscopic optimization step, the alternating direction method of multipliers (ADMM)-based convex optimization using the periodic method of moments (MoM) is implemented to deterministically obtain the surface parameters. Then, in the microscopic optimization step, the desired surface parameters are realized with physical unit cells using particle swarm optimization (PSO) integrated with machine learning (ML) surrogate models. Mutual coupling between the unit cells and edge effects that are of main concerns in the design of such finite surfaces are taken into account to minimize time-consuming and resource-demanding simulation-based optimization of the full array as well as avoiding using more manufacturing-intensive structures such as baffles (vias) \cite{Paul}. 

An electromagnetic-aware efficient method is used to generate training data for ML surrogate models. These models successfully and efficiently predict the magnitude as well as the phase of the scattering properties and replace the time-consuming full-wave simulations that are required for the optimization over the physical solution space. A new method to represent the 3-layer unit cells as the inputs of the surrogate models is implemented that reflects both the categorical and continuous nature of the solution space composed of different types of scatterers with various dimensions that are separated with different substrate thicknesses.

This paper is organized as follows. Section \ref{sec:II} details the macroscopic step to convert the high-level far-field constraints to surface parameters. In Section \ref{sec:III}, we explain the different parts of the microscopic step involving the ML-based surrogate models integrated in the particle swarm optimization to obtain the optimized physical EMMS. Two design examples to demonstrate the effectiveness of the proposed approach are presented in Section \ref{sec:IV}. Conclusions are drawn in Section \ref{sec:Conclusion}.

\section{Far-Field Constraints to Surface Parameters ${Z}_{se}$, ${Y}_{sm}$, and ${K}_{em}$ } \label{sec:II}

The first step of the inverse design process can be formulated as an optimization problem that searches for the  surface properties to satisfy local power conservation as well as the far-field constraints. After formulated as such, the problem can be solved deterministically. The details of this macroscopic optimization procedure have been presented \cite{SP}. The first step is to describe the EMMS as a fully homogenized structure using an integral equation approach implemented using the method of moments (MoM). 

\subsection{EMMS Model from the Method of Moments}
Starting with the generalized sheet transition conditions (GSTCs) for a bianisotropic EMMS,
\begin{subequations} 
\begin{align} 
\frac{1}{2}(\vec{E}_{t,1}+\vec{{E}}_{t,2}) &=\rttensortwo{{Z}}_{se}\cdot\vec{{J}}_{s} - \rttensortwo{{K}}_{em}\cdot(\hat{n}\times\vec{{M}}_{s})\\
\frac{1}{2}(\vec{{H}}_{t,1}+\vec{{H}}_{t,2}) &= \rttensortwo{{Y}}_{sm}\cdot\vec{{M}}_{s} + \rttensortwo{{K}}_{me}\cdot(\hat{n}\times\vec{{J}}_{s}).
\end{align}\label{GSTC_eqns}
\end{subequations}
These equations relate the tangential electric and magnetic fields on one side of the EMMS, $\vec{E}_{t,1}, \vec{H}_{t,1}$, to those on the other side, $\vec{E}_{t,2}, \vec{H}_{t,2}$, through the surface parameters and the surface currents. The surface parameters for this selected representation are the surface electric impedance ($\rttensortwo{{Z}}_{se}$), surface magnetic admittance ($\rttensortwo{{Y}}_{sm}$), and magneto-electric and electro-magnetic coupling coefficients ($\rttensortwo{K}_{em},\rttensortwo{{K}}_{me}$). The surface currents are the electric surface current density ($\vec{{J}}_{s}$) and magnetic surface current density ($\vec{{M}}_{s}$). 

\begin{figure}[!ht]
	\centering
  	\includegraphics[width=3in]{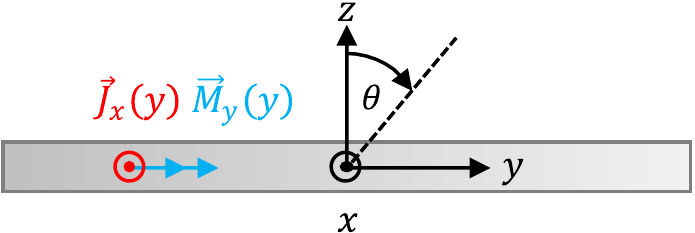}
  	\caption{Configuration of the homogenized EMMS model.}
  	\label{fig:HomogenizedEMMS}
\end{figure}
\vspace{0.5cm}

In this paper, we will use 2D TE-polarized examples. A diagram illustrating this configuration is shown in Figure \ref{fig:HomogenizedEMMS}, where the EMMS is periodic in the $x$-direction and varies nonuniformly in the $y$-direction. The EMMS will also be omega-type bianisotopic. As a result, the surface parameters in \eqref{GSTC_eqns} reduce to scalars as a function of position. In addition, we will only consider purely passive, lossless, and reciprocal EMMSs. This is because they are much easier to realize in practice. As a result,
\begin{subequations}
\begin{align}
\textrm{Re}({Z}_{se})=0\\
\textrm{Re}({Y}_{sm})=0\\
{K}_{em}={K}_{me}\\
\textrm{Im}({K}_{em})=0.
\end{align}
\end{subequations}
Since ${K}_{em}={K}_{me}$, we will simply use ${K}_{em}$ to denote the bianisotropic coupling terms. The details for the derivation of passive, lossless, and reciprocal omega-type bianisotropic GSTCs can be found in the work by Ataloglou \textit{et al} \cite{Ataloglou2021}.

Because we consider the EMMS within the model to have zero thickness, we can substitute the average fields in \eqref{GSTC_eqns} with the (known) incident and scattered fields. The scattered electric field due to an electric current density along a thin strip on the $y$-axis is given by
\begin{align}\label{Escat}
\vec{E}^{scat}_x(\vec{\rho}) 	= & -\frac{\omega\mu_0}{4}\int_{-w/2}^{w/2}\vec{J}_{x}(y')H_0^{(2)}(k|y-y'|)dy' 
\end{align}
where $w$ is the width of the thin strip and $\vec{\rho} = y\hat{y}$ is a position vector of the field point along the $y$-axis. Due to the fact that $\vec{M}_y$ is $y$-directed, it will not contribute to the scattered electric field in the $x$-direction. We can similarly derive the $y$-directed scattered magnetic field resulting from the magnetic current density as
\begin{align}\label{Hscat}
\begin{split}
\vec{H}^{scat}(\vec{\rho}) 	=& -\frac{1}{4\omega\mu_0}(k^2+\frac{\partial^2}{\partial y^2})\\ &\int_{-w/2}^{w/2}\vec{M}_{y}(y')H_0^{(2)}(k|y-y'|)dy'.
\end{split}
\end{align}
Similar to the scattered electric field, $\vec{{J}}_x$ is $x$-directed and will not contribute to the scattered magnetic field in the $y$-direction.

Using the MoM with pulse basis functions and point matching along the EMMS we can create linear system of equations \cite{SP},
\begin{subequations}
\begin{align}
\textit{\textbf{E}}^{inc} &= -\textit{\textbf{E}}^{scat}+j[\textbf{X}_{se}]\textit{\textbf{I}}^e-[\textbf{K}_{em}]\textit{\textbf{I}}^m\\
\textit{\textbf{H}}^{inc} &= -\textit{\textbf{H}}^{scat}+j[\textbf{B}_{sm}]\textit{\textbf{I}}^m+[\textbf{K}_{em}]\textit{\textbf{I}}^e\\
\textit{\textbf{E}}^{scat} &= -[\textbf{Z}_e]\textit{\textbf{I}}^e\\
\textit{\textbf{H}}^{scat} &= -[\textbf{Z}_m]\textit{\textbf{I}}^m,
\end{align}
\end{subequations}
where $\textit{\textbf{I}}^e$ and $\textit{\textbf{I}}^m$ are the basis coefficients for the electric and magnetic surface current, respectively. The elements of $[\textbf{Z}_e]$ and $[\textbf{Z}_m]$ are
\begin{subequations}
\begin{align}
z_e^{uv} &= \frac{\omega\mu_0}{4}\int_{(v-1)\Delta y}^{v\Delta y} H_0^{(2)}(k |y_u-y'|)dy'\\
\begin{split}
z_m^{uv} &= \frac{k^2}{8\omega\mu_0}\int_{(v-1)\Delta y}^{v\Delta y} H_2^{(2)}(k |y_u-y'|)  \\&+ H_0^{(2)}(k |y_u-y'|)dy',
\end{split}
\end{align}
\end{subequations}
where $u$ and $v$ are row and column indices, respectively.

In order to impose far-field constraints, we also need an equation to transform the discretized current densities and corresponding current coefficients  $\textit{\textbf{I}}^e$ and $\textit{\textbf{I}}^m$ to the far-field. We first examine the continuous form of the far-zone electric field at a distance $\rho$ resulting from electric and magnetic surface currents on the metasurface,
\begin{subequations}
\label{eq:Eff_cont}
\begin{align}
\begin{split}
\vec{E}_{ff}^e(\theta) &= - \frac{e^{-jk\rho}}{\sqrt{\rho}}\frac{\omega\mu_0}{4} \sqrt{\frac{2}{\pi k}}e^{j\frac{\pi}{4}} \\ & \int_{-w/2}^{w/2}\vec{J}_{x}(y')e^{jky'\sin\theta}dy'
\end{split}\\
\begin{split}
\vec{E}_{ff}^m(\theta) &= -\frac{e^{-jk\rho}}{\sqrt{\rho}}\frac{\omega\varepsilon_0 \eta_0}{4} \sqrt{\frac{2}{\pi k}}e^{j\frac{\pi}{4}} \\ &\int_{-w/2}^{w/2}\vec{M}_{y}(y')e^{jky'\sin\theta}\cos \theta dy'.
\end{split}
\end{align}
\end{subequations}
where $\theta$ is illustrated in Figure \ref{fig:HomogenizedEMMS}. In general, macroscopic optimization is more concerned with relative magnitudes of the far-zone pattern rather than the absolute value at a certain distance. As a result, we  focus on the far-zone fields irrespective of distance and frequency $\vec{E}_{0,ff}$. Now, using our discretized surface currents $\textit{\textbf{I}}^e$ and $\textit{\textbf{I}}^m$, we can rewrite \eqref{eq:Eff_cont} as a linear system of equations over $M$ angular far-zone samples and $N$ spatial samples along the surface as
\begin{subequations}\label{GIMatrix}
\begin{align}
\textit{\textbf{E}}_{ff}^e = & \textit{\textbf{E}}_{0,ff}^{e}\left(e^{-jk\rho}\sqrt{\frac{\lambda}{\rho}} \right)\\
\textit{\textbf{E}}_{0,ff}^{e} = & [\textbf{G}^e]\textit{\textbf{I}}^e\\
\textit{\textbf{E}}_{ff}^{m} = &  \textit{\textbf{E}}_{0,ff}^{m}\left(e^{-jk\rho}\sqrt{\frac{\lambda}{\rho}} \right)\\
\textit{\textbf{E}}_{0,ff}^{m} = & [\textbf{G}^m]\textit{\textbf{I}}^m,
\end{align}
\end{subequations}
where $[\textbf{G}^e] \in \mathbb{C}^{M\times N}$ and $[\textbf{G}^m] \in \mathbb{C}^{M\times N}$ are matrices expanding the discretized electric and magnetic surface current coefficients over their pulse basis functions and then transforming them to samples of far-zone electric field. These matrices have elements
\begin{subequations}\label{GIMatrixElements}
\begin{align}
{g}^{e,uv} = & -\frac{\omega\mu0}{4} \sqrt{\frac{2}{\pi k \lambda}}e^{j\frac{\pi}{4}}e^{jky_v\sin(\theta_u)}\Delta y \\
{g}^{m,uv} = & -\frac{\omega\epsilon_0\eta_0}{4} \sqrt{\frac{2}{\pi k\lambda}}e^{j\frac{\pi}{4}}e^{jky_v\sin(\theta_u)}\cos(\theta_u)\Delta y ,
\end{align}
\end{subequations}
where $u$ and $v$ are row and column indices, respectively. These elements were solved for by expanding $\vec{J}_{x}$ and $\vec{M}_{y}$ in (\ref{eq:Eff_cont}) over their pulse basis functions and integrating with the midpoint rule.

\subsection{Optimization Formulation}
We can now frame the EMMS macroscopic design as an optimization problem
\begin{subequations} \label{EMMS_basic_form}
\begin{align}
\minimize_{{\textit{\textbf{I}}}^e,{\textit{\textbf{I}}}^m,[{\textbf{X}}_{se}],[{\textbf{B}}_{sm}],[{\textbf{K}}_{em}]} &\quad f_0({\textit{\textbf{I}}}^e,{\textit{\textbf{I}}}^m,[{\textbf{X}}_{se}],[{\textbf{B}}_{sm}],[{\textbf{K}}_{em}])\\
\textrm{subject to}\quad\; \begin{split} & \quad {\textit{\textbf{E}}}^{inc} = [{\textbf{Z}}_e]{\textit{\textbf{I}}}^e+[{\textbf{X}}_{se}]{\textit{\textbf{I}}}^e \\ & \quad -[{\textbf{K}}_{em}]{\textit{\textbf{I}}}^m \end{split} \label{EMMS_E_const}\\
\begin{split}& \quad {\textit{\textbf{H}}}^{inc} = [{\textbf{Z}}_m]{\textit{\textbf{I}}}^m+[{\textbf{B}}_{sm}]{\textit{\textbf{I}}}^m\\ & \quad +[{\textbf{K}}_{em}]{\textit{\textbf{I}}}^e \end{split}\label{EMMS_H_const}\\
 \begin{split}& \quad f_i({\textit{\textbf{I}}}^e,{\textit{\textbf{I}}}^m,[{\textbf{X}}_{se}],[{\textbf{B}}_{sm}],[{\textbf{K}}_{em}]) \leq 0 \\ &\quad \quad i = 1,...,n,\end{split}
\end{align}
\end{subequations}
where $f_0$ and $f_i$ are some objective and constraint functions for a certain design goal. The optimization variables in this case are the surface electric and magnetic current coefficients ${\textit{\textbf{I}}}^e$ and ${\textit{\textbf{I}}}^m$ along with the surface electric reactance $[{\textbf{X}}_{se}]$, magnetic susceptance $[{\textbf{B}}_{sm}]$, and magneto-electric coupling $[{\textbf{K}}_{em}]$. Because we only optimize for passive and lossless surface parameters, our solution is passive and lossless by construction.

This optimization problem is non-convex due to the biaffine $[{\textbf{X}}_{se}]{\textit{\textbf{I}}}^e-[{\textbf{K}}_{em}]{\textit{\textbf{I}}}^m $ and $[{\textbf{B}}_{sm}]{\textit{\textbf{I}}}^m+[{\textbf{K}}_{em}]{\textit{\textbf{I}}}^e$ terms. Biaffine terms are created when two variables are multiplied together. Unfortunately, non-convex problems are notoriously hard to solve in polynomial time. Therefore, we cannot simply optimize for the surface currents to satisfy the objectives because that would likely yield active and/or lossy EMMSs. Optimizing the surface currents and parameters together yields a passive and lossless EMMS by construction. 

To solve this non-convex problem, we will relax it using ADMM \cite{Boyd2011}. ADMM is an algorithm that alternatingly minimizes the augmented Lagrangian of (\ref{EMMS_basic_form}) with respect to surface currents and parameters with each iteration. This is well suited to solving biaffine problems. Now that we can solve EMMS optimization problems in the form of \eqref{EMMS_basic_form}, we can devise far-field objective and constraint functions \cite{SP}.

\subsubsection{Main Beam Level}\label{MBLevel_section}
To try to force a beam to achieve a certain level (or get as close as possible), we can add a $\ell^2$-norm minimization term to the objective. For example, to force a beam at $\theta_0$ to a level $MB_{level}$ the objective function can take the form
\begin{align}\label{MBLevel}
\begin{split}
f_{MB}(\theta_0) =& \Vert [{\textbf{G}}^e](\theta_0){\textbf{\textit{I}}}^e+[{\textbf{G}}^m](\theta_0){\textbf{\textit{I}}}^m \\&+ {\textbf{E}}_{ff}^{inc}(\theta_0) - {MB}_{level}\Vert_2^2,
\end{split}
\end{align}
where ${\textbf{E}}_{ff}^{inc}(\theta_0)$ is the incident field across the extent of the surface in the far-field.

\subsubsection{Maximum Sidelobe Level} \label{SLL_section}
Enforcing a maximum permissible sidelobe over a certain set of angles can be done with the inequality constraint
\begin{align}
\vert {\textbf{G}}^e(SL){\textbf{\textit{I}}}^e+{\textbf{G}}^m(SL){\textbf{\textit{I}}}^m + {\textbf{E}}_{ff}^{inc}(SL) \vert \leq \tau + slack_{SL},
\end{align}
where we force the modulus of the total field over the sidelobe region $SL$ to be less than a certain sidelobe level $\tau$. The modulus is permissible here because it remains convex when used outside of an $\ell_2$-norm. We have included a slack variable $slack_{SL}$ so the following term must be added to the objective:
\begin{align} \label{eq:fSL}
f_{SL} = \Vert slack_{SL} \Vert^2_2
\end{align}
This is required in order to make the inequality active. 

\subsubsection{Surface Current Smoothness} \label{CurrentSmooth_section}
In order to allow for more feasible EMMS designs, we add a  surface current coefficient ${\textbf{\textit{I}}}^e$ and ${\textbf{\textit{I}}}^m$ smoothness constraint. This can be done with the inequality constraints
\begin{align}
\vert [\textbf{D}]{\textbf{\textit{I}}}^e\vert \leq D_{max}^e + slack_{D^e}\\
\vert [\textbf{D}]{\textbf{\textit{I}}}^m\vert \leq D_{max}^m + slack_{D^m},
\end{align}
where $[\textbf{D}]$ is the discrete second derivative matrix. Similar to (\ref{eq:fSL}), we have a function $f_D$ in the objective function to minimize the $\ell^2$-norm of the slack variables.

\subsubsection{Complete Formulation} \label{CompleteFormulation}
We can assemble a complete formulation if we fill in the objective function and inequality constraints of \eqref{EMMS_basic_form} with those described above. This can be written as

\begin{subequations} \label{EMMS_total_form}
\begin{align}
\minimize_{\substack{\textit{\textbf{I}}^e,\textit{\textbf{I}}^m,[\textbf{X}_{se}],[\textbf{B}_{sm}],[\textbf{K}_{em}],\\slack_{SL},slack_{D^e},slack_{D^m}}} \begin{split}&\quad\alpha_{MB}f_{MB}(MB)+\alpha_{NU}f_{NU}(NU)\\&\quad+\alpha_{SL}f_{SL}+\alpha_{D}f_{D}\end{split}\\
\textrm{subject to}  \quad\quad\;\;\;  \begin{split} & \quad {\textit{\textbf{E}}}^{inc} = [{\textbf{Z}}_e]{\textit{\textbf{I}}}^e\\ &\quad +[{\textbf{X}}_{se}]{\textit{\textbf{I}}}^e-[{\textbf{K}}_{em}]{\textit{\textbf{I}}}^m \end{split}\\
\begin{split}& \quad \beta({\textit{\textbf{H}}}^{inc} = [{\textbf{Z}}_m]{\textit{\textbf{I}}}^m\\&\quad+[{\textbf{B}}_{sm}]{\textit{\textbf{I}}}^m  +[{\textbf{K}}_{em}]{\textit{\textbf{I}}}^e)\end{split} \label{H_eqn}\\
\begin{split}&\quad\vert [{\textbf{G}}^e](SL){\textbf{\textit{I}}}^e+[{\textbf{G}}^m](SL){\textbf{\textit{I}}}^m \\ &\quad+ {\textbf{E}}_{ff}^{inc}(SL) \vert \leq \tau + slack_{SL}\end{split} \\
& \quad \vert [\textbf{D}]{\textbf{\textit{I}}}^e\vert \leq D_{max}^e + slack_{D^e} \\
& \quad \vert [\textbf{D}]{\textbf{\textit{I}}}^m\vert \leq D_{max}^m + slack_{D^m},
\end{align}
\end{subequations}
where $MB \in [0^\circ, 360^{\circ}]$ represents the set of main beam angles, $NU \in [0^\circ, 360^{\circ}]$ represents the set of null angles, $SL \in [0^\circ, 360^{\circ}]$ is the set of angles comprising the sidelobe region, and $\alpha_{MB},\alpha_{NULL},\alpha_{SL},\alpha_{D}$ are predetermined weights for different terms in the objective function. Due to different magnitudes of the electric and magnetic current MoM equations, a scaling term is needed. The $\beta$ term in (\ref{H_eqn}) is used to scale the magnetic current MoM equation. This is because when the augmented Lagrangian is formed, the equality constraints compete for minimization. Here, $\beta$ is experimentally determined to be $1000$. The weights are used if different portions of the optimization should be stressed more. For example, if the optimizer is having trouble satisfying the sidelobe level constraint, $\alpha_{SL}$ could be increased relative to the other weights. This is done on an experimental basis.

\section{From optimized Surface Properties to the EMMS Physical Structure} \label{sec:III}

To realize the optimized surface parameters that are obtained in Section \ref{sec:II} with actual physical EMMS, we consider the solution space of 3-layer bianisotropic unit cells. This solution space is composed of diverse scatterers with various dimensions that are separated with different substrate thicknesses and permittivity(ies). To obtain the optimum structure efficiently, we exploit and explore this solution space using a global optimizer integrated with machine-learning surrogate models. The surrogate models are trained on a limited set of training data to predict the unit cells' properties accurately and expedite the microscopic optimization process.     

\subsection{Generation of the Training Data } \label{sec:III.A}

\begin{figure}[!ht]
	\centering 
  	\includegraphics[width=3.6in]{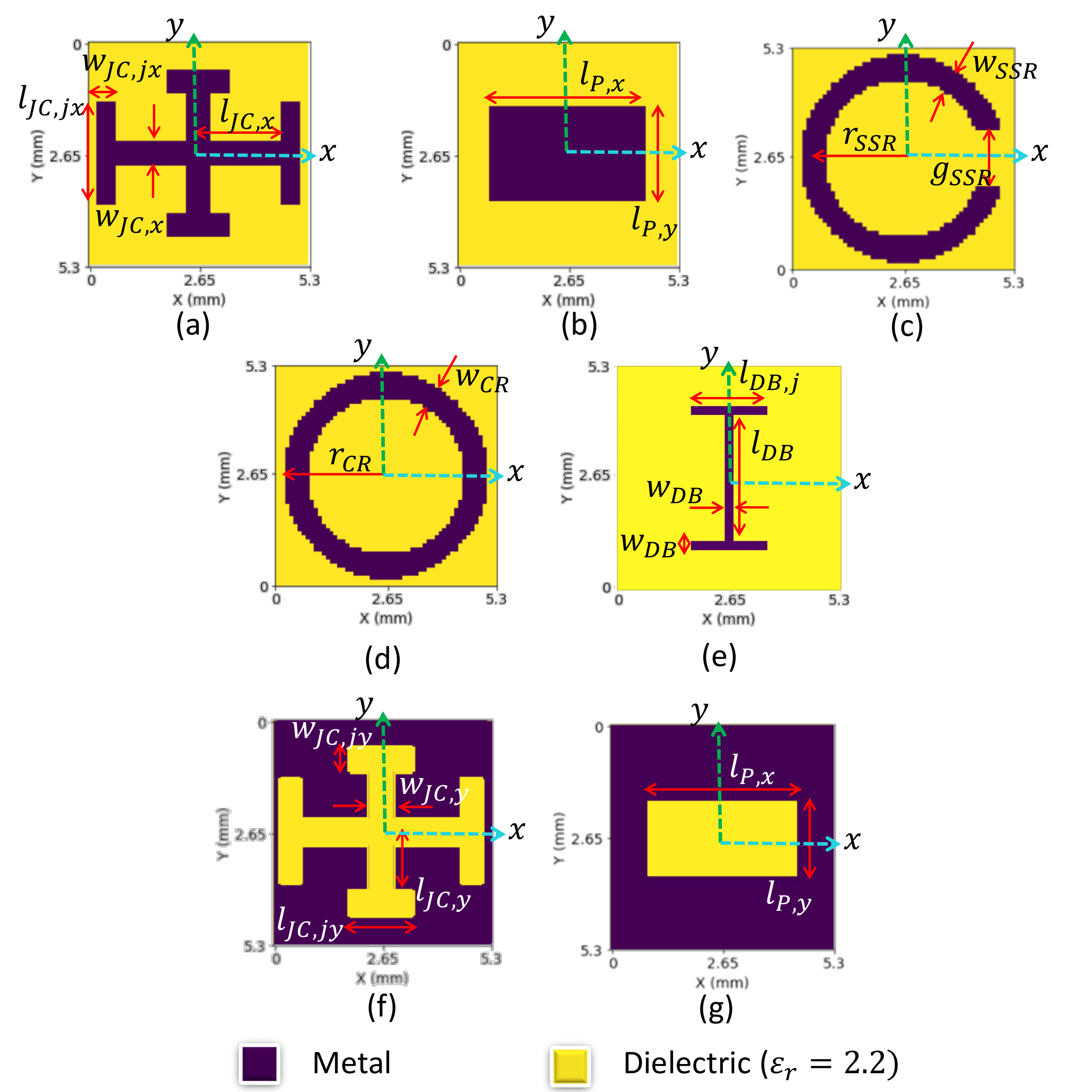}
  	\caption{Primitives used for training: (a) Jerusalem cross (JC), (b) rectangular patch (RP), (c) single split ring (SSR), (d) complete ring (CR), (e) dog bone (DB), (f) complementary Jerusalem cross (compJC), and (g) complementary rectangular patch (compRP).}
  	\label{shapes}
\end{figure}

The training data set is composed of 3-layer unit cells with a specific scatterer on each layer. Based on experience to curate the training data set, we choose certain scatterer shapes, called primitives, here. The shape of the primitives include the Jerusalem cross [Figure \ref{shapes} (a)], rectangular patch [Figure \ref{shapes} (b)], single split ring [Figure \ref{shapes} (c)], complete ring [Figure \ref{shapes} (d)], dog bone [Figure \ref{shapes} (e)]. The primitives shown in Figure \ref{shapes} (a)-(e) are used for the scatterers at the air-dielectric interface, i.e. the top and bottom layers. Often alternating inductive and capacitive behavior is required in bianisotropic EMMSs utilizing an odd number of layers. Therefore, we extend the range of scatterer geometries in the middle layer to include the complementary Jerusalem cross and complementary rectangular patch in Figure \ref{shapes} (f)-(g) as well. The values and ranges of the different dimensions of each primitive are listed in Table \ref{tab:table1}. Based on the range of each parameter, the numbesr of each primitive are also listed in Table \ref{tab:table1}.

The primitives are simulated from $1$ GHz to $19$ GHz with unit cell period of $5.3$ mm. Periodic boundary conditions on $x$- and $y$-sides are enforced. We numerically determine the generalized scattering matrix (GSM) containing not only the fundamental but also 10 other higher order modes, so that we can employ the cascading technique \cite{b24} to determine the 3-layer unit cell response \cite{b20}. This number of higher-order modes provides sufficiently accurate results to capture the interlayer coupling for the specified frequency range and unit cell period. Each scatterer is translated to meshes using Rao-Wilton-Glisson (RWG) basis functions and fed to our in-house periodic MoM-based simulation tool.

\captionsetup{skip=2pt}
\vspace{2mm}
\begin{table}[!]
\caption{Dimensions of the Primitives in Figure \ref{shapes}.}
\centering
\begin{tabularx}{\columnwidth}{|c|c|c|c|}
    \hline
     \textbf{Primitive} &  \textbf{Parameter} &  \textbf{Value (mm)} &  \textbf{Num. of Primitives} \\
    \hline \hline
    \multirow{4}{*}{} & $l_{JC,x/y}$ & $[2.2:0.2:4.0]$ &  \\
    \cline{2-3}
    JC & $l_{JC,cx/cy}$ & $l_{JC,x/y}-1.6$ mm & 100 (JCs) $\&$ \\
    \cline{2-3}
    $\&$ compJC & $w_{JC,x/y}$ & 0.4  & 100 (compJCs)\\
    \cline{2-3}
    & $w_{JC,cx/cy}$ & 0.45 & \\
    \hline
    \hline
    \multirow{2}{*}{RP } & $l_{P,x/y}$ & $[2.0:0.2:5.0]$ & 256 (RPs) $\&$  \\
    $\&$ comRP & & & 256 (compRPs)\\
    
    \hline
    \hline
    \multirow{3}{*}{SSR} & $r_{SSR}$ & $[1.4:0.2:2.6]$ & \multirow{3}{*}{202} \\
    \cline{2-3}
    & $g_{SSR}$ & $[0.3:0.2:1.5]$ & \\
	\cline{2-3}
    & $w_{SSR}$ & $[0.2:0.2:1.2]$ & \\
    \hline
    \hline
	\multirow{2}{*}{CR} & $r_{CR}$ & $[1.4:0.2:2.6]$ & \multirow{2}{*}{49} \\
    \cline{2-3}
    & $w_{CR}$ & $[0.1:0.2:1.3]$ & \\ 
    \hline
    \hline
    \multirow{3}{*}{DB} & $l_{DB}$ & $[2.2:0.2:4.8]$ & \multirow{3}{*}{280} \\
    \cline{2-3}
    & $l_{DB,c}$ & $[1.0:0.2:4.8]$ & \\
	\cline{2-3}
    & $w_{DB}$ & $0.2$ & \\
    \hline
    \hline

\end{tabularx}
\label{tab:table1}
\end{table}

The GSM calculation for each primitive is accelerated using a model-based parameter estimation (MBPE) method for the periodic MoM \cite{ZW}. This interpolation method provides accurate computation of the scatterers' GSMs over a broad frequency band by performing the calculation only at a few random frequencies in the band. Patch-like scatterers can be simulated using MBPE by solving MoM with the electric-field integral equation (EFIE) \cite{ZW}. To handle general cases including complementary scatterers, an extension on the existed MBPE method has been developed for magnetic-field integral equation (MFIE)-based MoM. Therefore, we are capable to efficiently generate a large data set required for the ML analysis including all types of periodic structures.

3-layer EMMS training samples are generated by randomly selecting three different GSMs (to implement bianisotropy) and cascading them with different dielectric thicknesses including $0.254$, $0.508$, $0.787$, and $1.575$ mm. About $70000$ samples out of about $2$ billion possible combinations are generated in this way. This electromagnetic-aware approach to generate the training data set is more efficient than the traditional approach of simulating the 3-layer unit cells. The training data generation was parallelized using the multiprocessing package from Python. The training set is denoted by $X_t$ and the training subsets for each substrate thickness are denoted by $X_{t,h}$ here. 

\subsection{The EMMS Unit Cell Representation} \label{sec:III.B}

To explore the design space, new 3-layer combinations of the primitives shown in Figure \ref{shapes} need to be evaluated beyond those generated in the 70000 samples described in Section \ref{sec:III.A}. For that, we exploit the samples in the training data set to train deep learning neural networks to provide fast predictions of the scattering parameters of the candidate under test. 

The solution space contains different scatterer primitives, standard substrate thicknesses, and continuous scatterer dimensions. Therefore, we describe each EMMS unit cell with a combination of categorical and continuous variables. The categorical parts are used to describe which scatterer primitive for each layer and what substrate thickness are used in the unit cell while the continuous variables describe the relevant dimensions of the scatterers for each layer.

\begin{figure}[!ht]
	\centering
  	\includegraphics[width=3.6in]{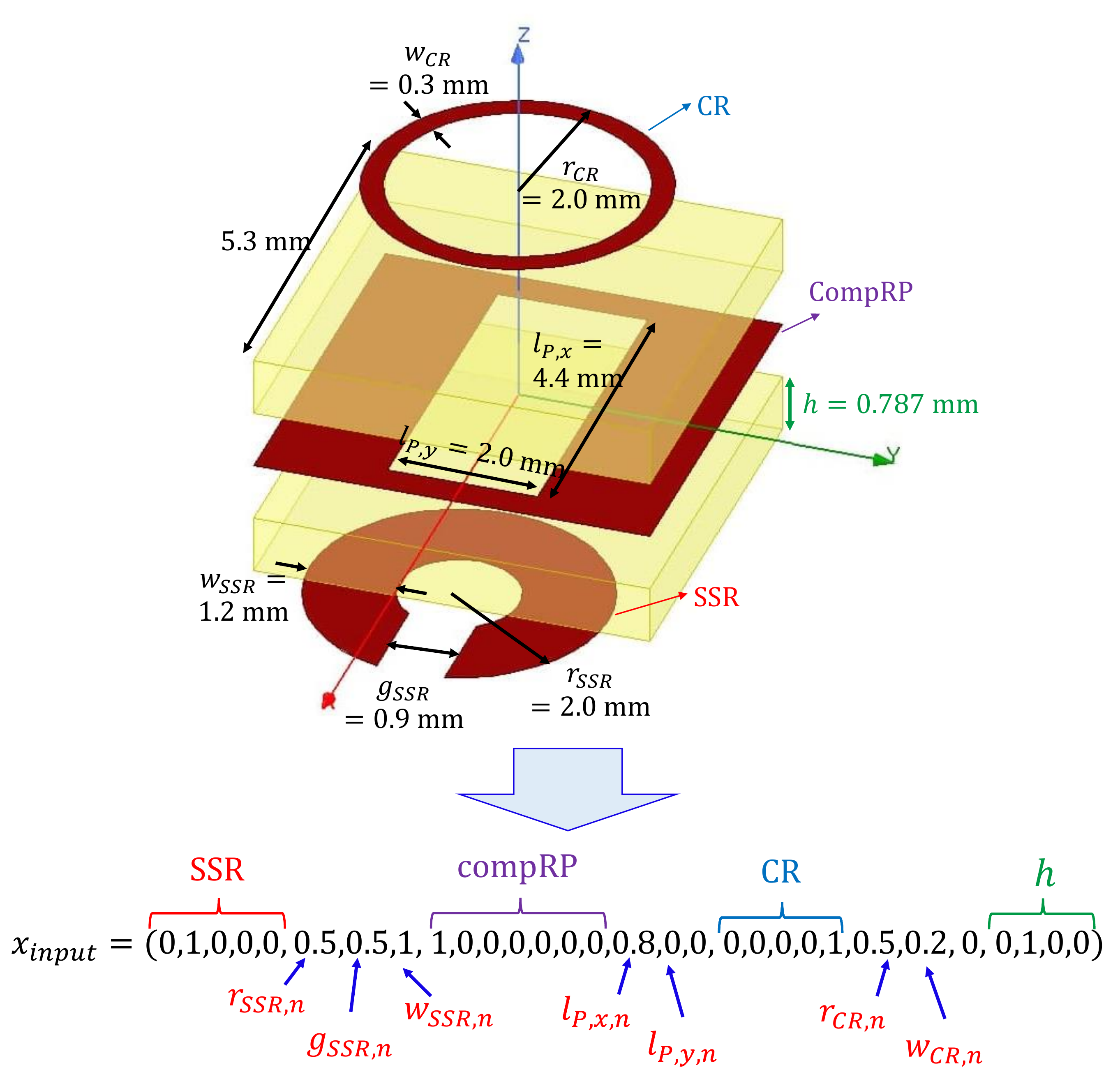}
  	\caption{An example of how the physical structure of each unit cell is translated to a 30-variable input for the surrogate models.   }
  	\label{fig3_2}
\end{figure}

Table \ref{tab:table2} shows the code and the categorical variables given to each scatterer shape and the substrate thickness. It is worth noting that five types of the scatterer primitives shown in Figure \ref{shapes} (a)-(e) can be used for the top and the bottom layers, while all the shapes in Figure \ref{shapes} (a)-(g) can construct the middle layer of a unit cell. Therefore, five digit and seven digit variables are used to describe the top/bottom-layer and the middle-layer scatterers, respectively. Only the varying dimensions of the scatterers, as shown in Figure \ref{shapes}, are represented with three values between $0$ and $1$, where $0$  and $1$ represent the minimum and maximum of each parameter in Table \ref{tab:table1}, respectively. This is an important adjustment to emphasize on the impact of each physical parameter in the scatterer design. The order of the normalized dimensions are based on their appearance in Table \ref{tab:table1} for each scatterer. It is worth noting that the dependent and fixed dimensions of the scatterers, e.g. $l_{JC,cx/cy}$ and $w_{JC,cx/cy}$ for the Jerusalem cross, are not included in this representation. For all the shapes except the single split ring (SSR) shown in Figure \ref{shapes} (c), the third value is fixed to $0$ since only two physical parameters are to be tuned. This is done to keep the length of the representation for all the unit cell combinations fixed. Therefore, the top and bottom layer scatteres are described by $8$ variables each and the middle layer is described by $10$ variables. Adding the substrate thickness representation, this results in a total of $30$ variables to describe each sample 3-layer unit cell in the training data. The mapping from the physical unit cell space to the $30$-dimensional solution space is shown through an example in Figure \ref{fig3_2}. This type of representation greatly helps the surrogate models to understand both the categorical and continuous aspects of the input data, i.e. physical structures of the EMMSs. We elaborate on this further after introducing the surrogate models in Section \ref{sec:III.C}. 

\vspace{1cm}

\begin{table}[ht]

\caption{The code and categorical variables assigned to each scatter shape and each substrate thickness.}
\centering
\begin{tabularx}{\columnwidth}{|c|c|X|}
    \hline
    \textbf{Type} &  \textbf{Code} &  \textbf{Categorical Variables}  \\
    \hline \hline 
    \rowcolor[HTML]{C0C0C0} \multicolumn{3}{c}{\textbf{Scatterer Type}} \\
    \hline
    \hline \xrowht[()]{5pt}

    \multirow{2}{*}{CR} &  \multirow{2}{*}{$0$} & \text{If on top/bottom layer:} $(0,0,0,0,1)$ \\
   \cline{3-3} \xrowht[()]{5pt}
    & & \text{If on middle layer:} $(0,0,0,0,0,0,1)$ \\ 
    \hline \xrowht[()]{5pt}
    
     \multirow{2}{*}{JC} &  \multirow{2}{*}{$1$} & \text{If on top/bottom layer:} $(0,0,0,1,0)$ \\
   \cline{3-3} \xrowht[()]{5pt}
    & & \text{If on middle layer:} $(0,0,0,0,0,1,0)$ \\
    \hline \xrowht[()]{5pt}
    
      \multirow{2}{*}{RP} &  \multirow{2}{*}{$2$} & \text{If on top/bottom layer:} $(0,0,1,0,0)$ \\
   \cline{3-3} \xrowht[()]{5pt}
    & & \text{If on middle layer:} $(0,0,0,0,1,0,0)$ \\
    \hline \xrowht[()]{5pt}
     \multirow{2}{*}{SSR} &  \multirow{2}{*}{$3$} & \text{If on top/bottom layer:} $(0,1,0,0,0)$ \\
   \cline{3-3} \xrowht[()]{5pt}
    & & \text{If on middle layer:} $(0,0,0,1,0,0,0)$ \\
    \hline \xrowht[()]{5pt}
     \multirow{2}{*}{DB} &  \multirow{2}{*}{$4$} & \text{If on top/bottom layer:} $(1,0,0,0,0)$ \\
   \cline{3-3} \xrowht[()]{5pt}
    & & \text{If on middle layer:} $(0,0,1,0,0,0,0)$ \\
    \hline \xrowht[()]{5pt}
     \multirow{2}{*}{compJC} &  \multirow{2}{*}{$5$} & \text{If on top/bottom layer:} $N/A$ \\
   \cline{3-3} \xrowht[()]{5pt}
    & & \text{If on middle layer:} $(0,1,0,0,0,0,0)$ \\
    \hline \xrowht[()]{5pt}
    
    \multirow{2}{*}{compRP} &  \multirow{2}{*}{$6$} & \text{If on top/bottom layer:} $N/A$ \\
   \cline{3-3} \xrowht[()]{5pt}
    & & \text{If on middle layer:} $(1,0,0,0,0,0,0)$ \\
    \hline 
    \hline
	\rowcolor[HTML]{C0C0C0} \multicolumn{3}{c}{\textbf{Substrate thickness}} \\
    \hline
    \hline \xrowht[()]{5pt}
    $0.254$ mm & $0$ & $(0,0,0,1)$ \\ 
    \hline \xrowht[()]{5pt}
     $0.508$ mm & $1$ & $(0,0,1,0)$ \\ 
    \hline \xrowht[()]{5pt}
    $0.787$ mm & $2$ & $(0,1,0,0)$ \\ 
    \hline \xrowht[()]{5pt}
    $1.575$ mm & $3$ & $(1,0,0,0)$ \\ 
    \hline
\end{tabularx}
\label{tab:table2}
\end{table}

\subsection{Surrogate Models} \label{sec:III.C}

We use the $30$-variable representation, $x_{input}$, for the physical structure of each unit cell along with the normalized frequency point as the inputs to deep neural networks (DNNs), acting as surrogate models, to predict the magnitudes and phases of the dispersive scattering parameters in an accelerated. To find the optimized EMMS design, a loss function that quantifies the difference between the desired surface properties and the surface properties offered by the example under test needs to be evaluated. Therefore, for each unit cell that is being examined, the surface properties need to be obtained in the optimization loop. Traditionally, full-wave solvers are used to perform this part. However, due to the large amount of time and resources they need, this step is the bottleneck of the optimization through a large and high-dimensional solution space such as the one of a bianisotropic EMMS. To accelerate this part, we replace the full-wave solver with surrogate models.

Provided with enough training examples, DNNs have proved to be efficient tools to learn the underlying patterns in large data sets and use this patterns to provide fast predictions for the features of the new inputs that they have never encountered before. These deep neural networks can then be integrated into optimization cycles as surrogate models to accelerate the search for the optimized features. 

Here, we choose to predict the scattering parameters of the EMMS unit cell instead of ${Z}_{se}$, ${Y}_{sm}$, and ${K}_{em}$. This is because the magnitudes and phases of the scattering parameters are intrinsically bounded and less prone to drastic changes at the scatterer's resonant frequencies. Without the loss of generality, we focus on developing the surrogate models for TE-polarized scattering parameters. The DNNs used to predict the magnitudes and phases of the scattering parameters are designated \textit{mag}-DNN and \textit{phase}-DNN, respectively. The discussions here can be applied to predict TM-scattering parameters in the same models of the TE-scattering parameters or in separate models. However, building the right DNN that can accurately predict the required features necessitates setting different hyperparameters including the number of the hidden layers and the number of the neurons in each layer. More importantly, one can consider the inputs to the DNN the key factor in its success. 

\begin{figure}[!ht]
	\centering
  	\includegraphics[width=2.5in]{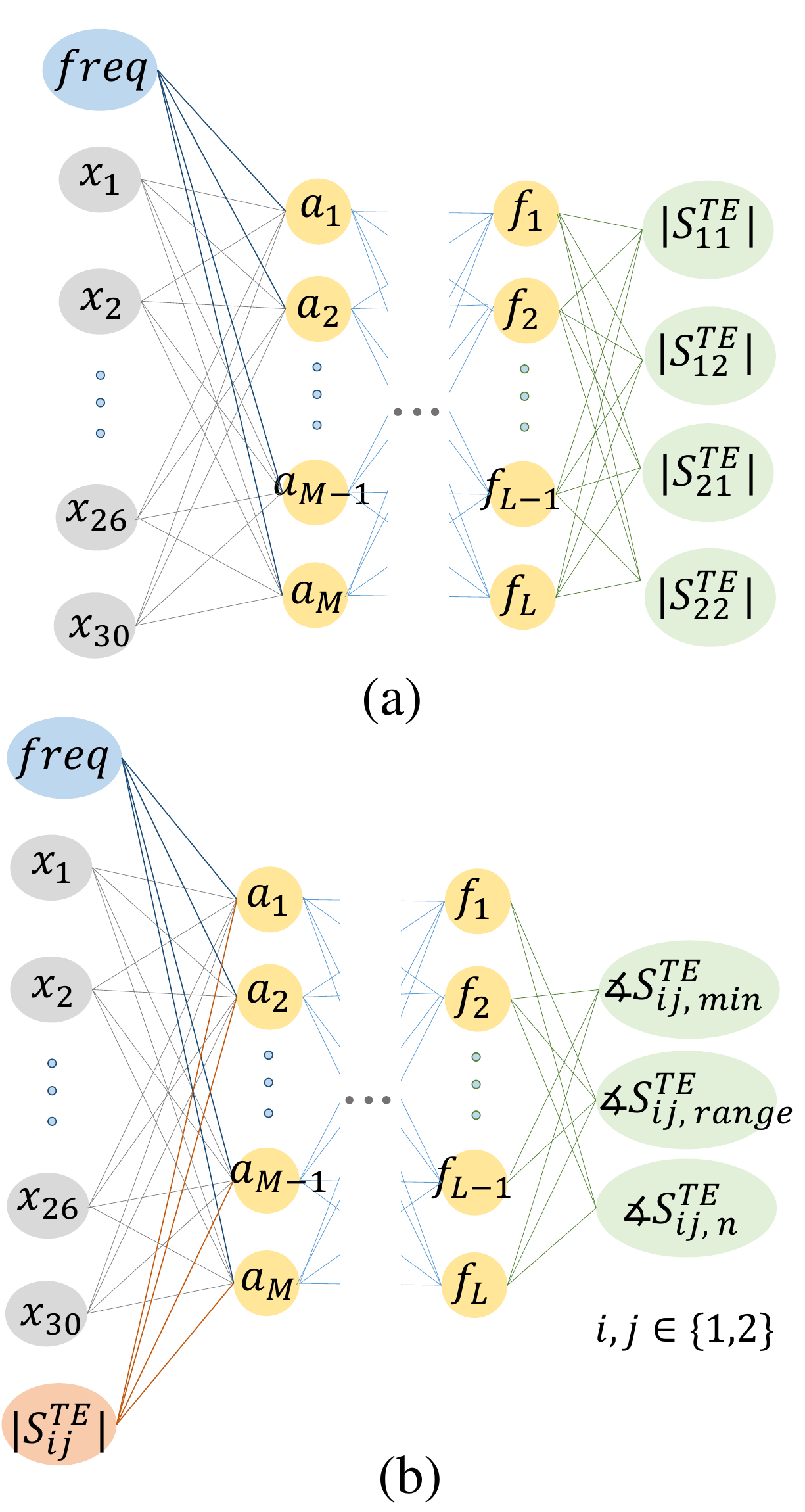}
  	\caption{Deep neural networks of the surrogate models to predict (a) the magnitude of the TE-scattering parameters (\textit{mag}-DNN), and (b) the phase of the $S_{12}^{TE}$ (\textit{phase}-DNN). }
  	\label{fig3_4}
\end{figure}

Figure \ref{fig3_4} (a) shows the \textit{mag}-DNN. One might consider using only $x_{input}$ to predict a vector of scattering parameters, where elements of the vector represent the predicted scattering parameter at different frequency points in the band. However, our tests reveal that it is simpler and more successful to train an average-size DNN to predict the scattering parameters at each frequency based on the frequency point input to the DNN. This choice greatly helps the DNN to make accurate predictions especially at the resonant frequency(ies) of the unit cell. 

We implemented the DNNs using TensorFlow-backend Keras libraries \cite{keras} in Python. The trained \textit{mag}-DNN has six hidden layers with $100$, $500$, $1000$, $1000$, $500$, $100$ neurons with ReLU activation functions, from the most shallow to the deepest layers, respectively. The output layer has four neurons for the magnitude of each TE-scattering parameter with sigmoid activation function due to the bounded values between $0$ and $1$. The DNN is trained with a batch size of $2048$ using the ADAM optimizer \cite{Adam} and backpropagation method \cite{b23} on $85\%$ of the training set and tested on the remaining $15\%$. The loss function used to train the \textit{mag}-DNN is based on the mean squared error (MSE) between the actual and predicted values as described by

\begin{subequations}
\label{eq:3_4}
\begin{dmath}
L_{\textit{mag}-DNN} = \sum \{\textrm{MSE}(|S_{11}|)+\textrm{MSE}(|S_{12}|)+ \textrm{MSE}(|S_{21}|)+\textrm{MSE}(|S_{22}|)\},   \label{eq:3_4c} 
\end{dmath}
\text{where}
\begin{dmath}
\textrm{MSE}(|S_{ij}|) = |||S_{ij,actual}^{TE}| -|S_{ij,pred}^{TE}|||_2^2 ; \quad i,j \in {1,2}.         \label{eq:3_4a}
\end{dmath}
\end{subequations}

\vspace{1cm}
Unlike the magnitudes of the scattering parameters, the scattering parameter phases are not limited to values between $0$ and $1$. In fact, different unit cells have a different phase range in the frequency band considered. Moreover, there are $180^\circ$- and $360^\circ$-phase jump(s) based on the order of unit cell's responses. These features make the accurate predictions for the scattering parameter phases more challenging than the ones for the magnitudes because DNNs are better at predicting bounded values within a specific range. Therefore, we consider the following points to implement a DNN to predict each TE-scattering parameter's phase at different frequencies within the band.

\begin{enumerate}
  \item Adding the magnitude of the scattering parameter to the input of the \textit{phase}-DNN: It is known that when the magnitude of the scattering parameter is zero, its phase experiences a jump. Therefore, to predict the frequency of these phase jumps, we use the magnitude information as an auxiliary input to the DNN besides the normalized frequency point and the $30$-variable  physical representation of the EMMS. This greatly improves the accuracy of the predictions made by the \textit{phase}-DNN.    
  \item Predicting $\measuredangle S_{ij, min}^{TE}$, $\measuredangle S_{ij, range}^{TE}$, and $\measuredangle S_{ij, n}^{TE}$ instead of $\measuredangle S_{ij}^{TE}$ , where $i,j \in \{1,2\}$ and
\begin{subequations}
\label{eq:3_5}
\begin{align}
\measuredangle S_{ij, min}^{TE} &= \min(\textrm{unwrap}\{\measuredangle S_{ij}^{TE}\}),         \label{eq:3_5a} \\
\measuredangle S_{ij, range}^{TE} &=  \max(\textrm{unwrap}\{\measuredangle S_{ij}^{TE}\} - \measuredangle S_{12, min}^{TE}) , \label{eq:3_5b} \\
\measuredangle S_{ij, n}^{TE}&= \frac{\textrm{unwrap}\{\measuredangle S_{ij}^{TE}\} - \measuredangle S_{ij, min}^{TE}}{\measuredangle S_{ij, range}^{TE}}.  \label{eq:3_5c}\\
 \measuredangle S_{ij}^{TE}&= \measuredangle S_{ij, n}^{TE} \times \measuredangle S_{ij, range}^{TE} +\measuredangle S_{ij, min}^{TE}  
 \label{eq:3_5d}
\end{align}
\end{subequations}
Instead of predicting $\measuredangle S_{ij}^{TE}$ that can range from $\measuredangle S_{ij, min}^{TE}$ and $\measuredangle S_{ij, range}^{TE}$ at each frequency, \textit{phase}-DNN predicts $\measuredangle S_{ij, min}^{TE}$ and $\measuredangle S_{ij, range}^{TE}$ that are constant for a certain unit cell over the frequency points and $\measuredangle S_{ij, n}^{TE}$ that is bounded between $0$ and $1$ for all unit cells.
\end{enumerate}
Both of these adjustments greatly help in understanding the relation between the inputs and outputs by \textit{phase}-DNN.

\textit{phase}-DNN has six hidden layers with $100$, $500$, $2000$, $2000$, $500$, $100$ neurons with ReLU activation functions, from the most shallow to the deepest layers, respectively. The output layer has three neurons for  $\measuredangle S_{ij, min}^{TE}$, $\measuredangle S_{ij, range}^{TE}$, and $\measuredangle S_{ij, n}^{TE}$. One of these output neurons that is intended to provide prediction of $\measuredangle S_{ij, n}^{TE}$ has a sigmoid activation function and the other two do not employ any activation functions. The \textit{phase}-DNN is trained in a similar way to the \textit{mag}-DNN with the following loss function to be minimized:
\begin{dmath}
L_{\textit{phase}-DNN} = \sum \{\textrm{MSE}(\measuredangle S_{ij,min}^{TE})+\textrm{MSE}(\measuredangle S_{ij,range}^{TE})+ \textrm{MSE}(\measuredangle S_{ij,n}^{TE})\}.   \label{eq:3_6} 
\end{dmath}
Firstly, one \textit{phase}-DNN is trained to predict phases of $S_{12}^{TE}$ of different unit cells at different frequency points. Then, transfer learning method \cite{TL} is employed to train similar neural networks to $S_{12}^{TE}$'s \textit{phase}-DNN for the predictions of the rest of the scattering parameter phases, e.g. phase of $S_{11}^{TE}$. In this method, instead of beginning the training of the new DNN from random weights, the DNN is initialized with the weights of the pretrained $S_{12}^{TE}$'s \textit{phase}-DNN. This approach requires much fewer number of epochs to minimize the prediction error, which considerably reduces the training time.

\begin{figure}[!ht]
	\centering
  	\includegraphics[width=3in]{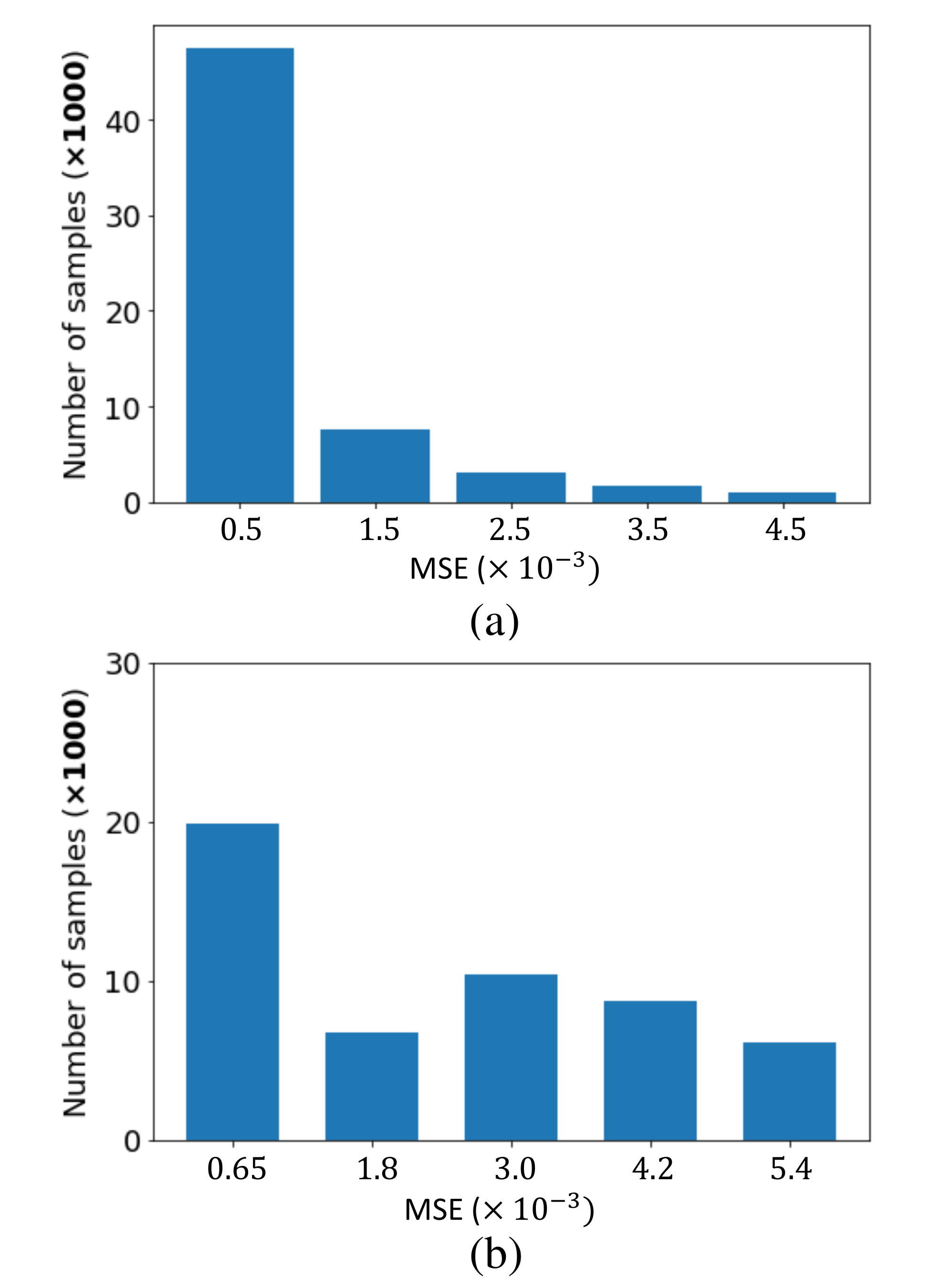}
  	\caption{MSE distribution over all the training samples for (a) the \textit{mag}-DNN in Figure \ref{fig3_4} (a), and (b) the \textit{phase}-DNN in Figure \ref{fig3_4} (b) trained to predict $\measuredangle S_{12}^{TE}$. }
  	\label{fig3_5}
\end{figure}

Figures \ref{fig3_5} (a) and (b) show the MSE distribution over the training data set. It can be seen that most of the samples are predicted with an error of around $5.0 \times 10^{-4}$ and $6.5 \times 10^{-3}$ rad by the \textit{mag}-DNN and \textit{phase}-DNN, respectively. The average error over the training set for \textit{mag}-DNN and \textit{phase}-DNN are $8.8 \times 10^{-4}$ and $2.4 \times 10^{-3}$ rad, respectively. 

\subsection{Optimization Over the Solution Space} \label{sec:III.D}

Practical implementation of the EMMS dictates that a certain substrate thickness must be chosen for all the unit cells that realize the desired surface parameters. To find which substrate thickness must be used, the following fast approach is implemented. For each EMMS unit cell, the minimum error between the desired scattering parameters and the training samples of each substrate thickness, $X_{t,h}$, are calculated as 
  
 \begin{dmath}\label{eq:Lh} 
 L({h}) = \sum_{i=1}^{i=N}{ \textrm{min} (||S_{11,desired,i} - S_{11,X_{t,h}}||_2^2 + ||S_{12,desired,i} - S_{12,X_{t,h}}||_2^2}),
 \end{dmath} 
 where $N$ is the total number of unit cells in the EMMS. $L(h)$ is calculated for each of the $h \in \{0.254, 0.508, 0.787, 1.525 \}$ mm and the thickness that corresponds with the minimum value of $L(h)$ is chosen as the optimized sub-solution space to be explored.

After finding which sub-domain provides the maximum number of matches to the desired scattering properties, the optimization through this space is performed as follows. To obtain the physical structure of the $i$-th unit cell of the EMMS, particle swarm optimization (PSO) is used. The $i$-th desired scattering parameters converted from the optimized $i$-th $Z_{se}, Y_{sm},$ and $K_{em}$ obtained from Section \ref{sec:II} are defined as the targets in the loss function of the PSO. This loss function is defined as $L_{PSO,i} = |[S]_{desired,i} - [S]_{pred}|$, where $[S]_{pred}$ is efficiently computed using the surrogate models developed in Section \ref{sec:III.C}. 

The swarm of the PSO includes $P$ particles and it is run for $I$ iterations. The positions of each particle are denoted by $x_n$, where $n \in [1,P]$ and $x$ denotes the EMMS unit cell representation under test that has $30$ variables.   $x_n$ is updated at the $(m+1)$th iteration based on \cite{PSO}
\begin{equation} 
	{x_n(m+1) = x_n(m) + v_n(m+1)},
\end{equation}
{where }
\begin{equation}
{x_n(m+1) = w \times v_{n,i}(m) + c_1 \times (p_n - x_n) + c_2 \times (p_g - x_n).}
\end{equation}
$p_n$ is the particle's historically best position and ${p_g}$ is the swarm's best position regardless of which particle had found it. $c_1$  and $c_2$ are the cognitive and social parameters, respectively. They control the particle's behavior given two choices: (1) to follow its personal best or (2) follow the swarm's global best position. Overall, this determines if the swarm is explorative or exploitative in nature. In addition, a parameter $w$ controls the inertia of the swarm's movement. The performance of the PSO \cite{PSO} is controlled by the choices of $P$, $I$, $c_1$, $c_2$, and $w$. Here, these parameters are empirically selected to obtain the best results. The PSO is implemented using the PySwarms library in Python \cite{pyswarms}.

\section{Design Examples}\label{sec:IV}

In this section, two examples are provided with the far-field constraints including main beam(s) levels, sidelobe level, and null position applied on the transmitting side of the EMMS. The design of both surfaces is carried out at $9.4$ GHz where the unit cell's period, i.e. $5.3$ mm, is $\lambda_0 /6$, where $\lambda_0$ is the free space wavelength. The EMMS has a size of $D = 15 \lambda_0$ composed of $90$ unit cells. The EMMS unit cells are located at $z= 0$ mm and distributed in the interval of $y \in  [-7.5 \lambda_0, 7.5 \lambda_0]$. Since the field on the transmitted side is mostly impacted by the magnitude and phase of the $S_{12}^{TE}$, we modify the PSO loss function at $9.4$ GHz to 
  
  \begin{dmath} \label{eq:LPSOmod}
 L_{PSO,i} = ||S_{11,desired,i} - S_{11,pred}||_2^2 + |||S_{12,desired,i}| - |S_{12,pred}|||_2^2.
 \end{dmath}

\begin{figure}[!ht]
	\centering
  	\includegraphics[width=3.6in]{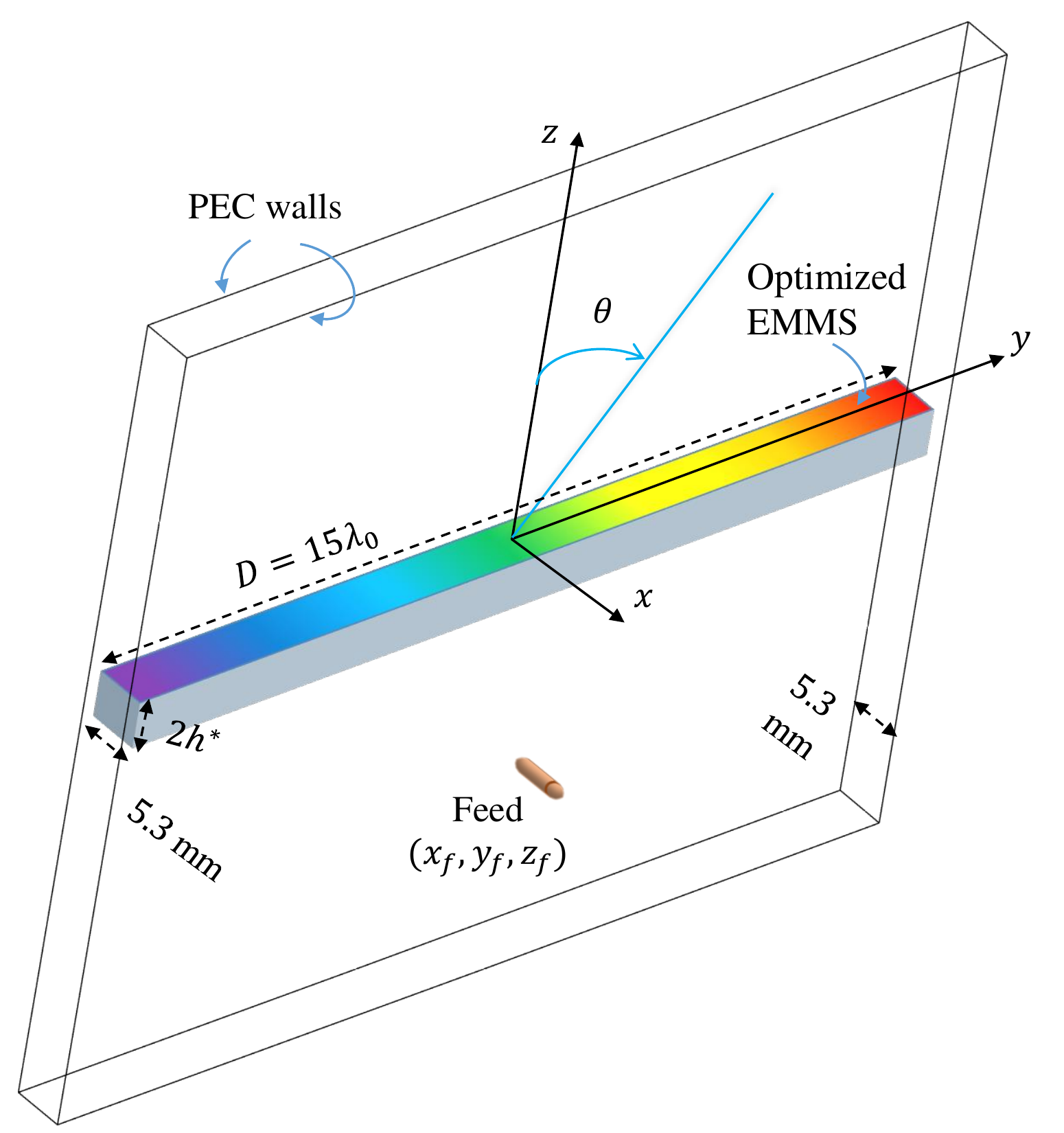}
  	\caption{Simulation setup of the optimized EMMS. }
  	\label{fig:sim_setup}
\end{figure}

To analyze the optimized EMMS for each case, we use scripts that read the file containing the primitives and dimensions of the EMMS scatterers and generate their model in Ansys HFSS. Once the 1D array of unit cells is generated, perfect boundary conditions (PEC) are assigned at $x=-2.65$ mm and $x=2.65$ mm to replicate a periodic array in that direction and achieve a 2D simulation. This setup is shown in Figure \ref{fig:sim_setup}.

\subsection{Example 1: $45^\circ$-Refraction Optimization}
In this example,  the $15 \lambda_0$ EMMS is excited with two line sources that are located at $x_f = 0$ mm, $y_f = \pm D/8$, and $z_f=- D/4$. The far-field constraints are defined to transform the excitation field to a beam directed at $\theta=45^ \circ$ with sidelobe levels $12$ dB below relative to the main beam for angles $ \theta \in [0,38] ^\circ$ and $10$ dB below the main beam for the remainder of the angles. Moreover, two null regions in the front, $\theta = 0 ^\circ$, and back, $\theta \in \{179,180,181\} ^\circ$, of the EMMS are defined. The details of this optimization problem are listed in Table \ref{table:45deg_table} and  \eqref{eq:45degProblem}.

\begin{table}[!t]
\caption{$45^\circ$-Refraction optimization parameters}

\label{table:45deg_table}
\centering
\begin{tabular}{|c|c|}
\hline
 Parameter &  Value \\
\hline
 Incident  & $\frac{1}{\sqrt{2r_1}}e^{-jkr_1}+\frac{1}{\sqrt{2r_2}}e^{-jkr_2}$ \\
 Field  ($E^{inc}$)  & $r_1 = \sqrt{(y-D/8)^2+(D/4)^2 }$ \\ 
& $r_2 =  \sqrt{(y+D/8)^2+(D/4)^2 }$\\
\hline
 Angular Sampling Points ($M$) & $361$ \\
\hline
 Spatial Sampling Points ($N$) & $90$ \\
\hline
 Max Iterations & $295$ \\
\hline
 Initial & $([{\textbf{X}}_{se}]^0,[{\textbf{B}}_{sm}]^0,[{\textbf{K}}_{em}]^0) = \textbf{0}, $\\
 Conditions & $\rho = 10,(\bm{\mu}_{\textbf{Z}_e}^0, \bm{\mu}_{\textbf{Z}_m}^0) = \textbf{0}$ \\
 \hline
 Main Lobe Angle ($ MB$) & $ \theta = 45^{\circ}$ \\
 \hline
 Main Lobe Level ($MB_{level}$)  & $ 16.8+j16.8$ \\
\hline
 Sidelobe  & $ \lbrace  7.4272 ,0^{\circ}\leq \theta\leq 38^{\circ}\rbrace ,$\\
 Level ($\bm{\tau}$)& $ \lbrace  5.9679,  51^{\circ}\leq \theta\leq 180^{\circ}\rbrace ,$ \\
 and Angles ($SL$)  &  $ \lbrace 7.4272,  -180^{\circ}\leq \theta\leq -90^{\circ}\rbrace , $\\
 & $  \lbrace 7.4272,  -90^{\circ}\leq \theta\leq 0^{\circ}\rbrace$\\
\hline
 Null Angles ($NU$)& $\theta=\left\lbrace 0^{\circ},179^{\circ},180^{\circ},181^{\circ}\right\rbrace$  \\
\hline
\end{tabular}
\end{table}

\begin{subequations} \label{eq:45degProblem}
\begin{align}
\minimize_{\substack{\textit{\textbf{I}}^e,\textit{\textbf{I}}^m,[\textbf{X}_{se}],[\textbf{B}_{sm}],[\textbf{K}_{em}],\\slack_{D^e},slack_{D^m}}} \begin{split}&\quad 10000f_{MB}(MB)+\\ & \quad 10000f_{NU}(NU)+f_{D}\end{split}\\
\textrm{subject to}  \quad\quad\;   & \quad {\textit{\textbf{E}}}^{inc} = [{\textbf{Z}}_e]{\textit{\textbf{I}}}^e+[{\textbf{X}}_{se}]{\textit{\textbf{I}}}^e-\\ & \quad[{\textbf{K}}_{em}]{\textit{\textbf{I}}}^m \\
\begin{split}& \quad 1000({\textit{\textbf{H}}}^{inc} = [{\textbf{Z}}_m]{\textit{\textbf{I}}}^m+\\ & \quad[{\textbf{B}}_{sm}]{\textit{\textbf{I}}}^m +[{\textbf{K}}_{em}]{\textit{\textbf{I}}}^e)\end{split} \label{H_eqn}\\
\begin{split}&\quad\vert [{\textbf{G}}^e](SL){\textbf{\textit{I}}}^e+[{\textbf{G}}^m](SL){\textbf{\textit{I}}}^m \\ &\quad+ {\textbf{E}}_{ff}^{inc}(SL) \vert \leq \bm{\tau} \end{split} \\
& \quad \vert [\textbf{D}]{\textbf{\textit{I}}}^e\vert \leq \textbf{0.1} + slack_{D^e} \\
& \quad \vert [\textbf{D}]{\textbf{\textit{I}}}^m\vert \leq \textbf{25}+ slack_{D^m}
\end{align}
\end{subequations}

\begin{figure}[!h]
	\centering
  	\includegraphics[width=3.5in]{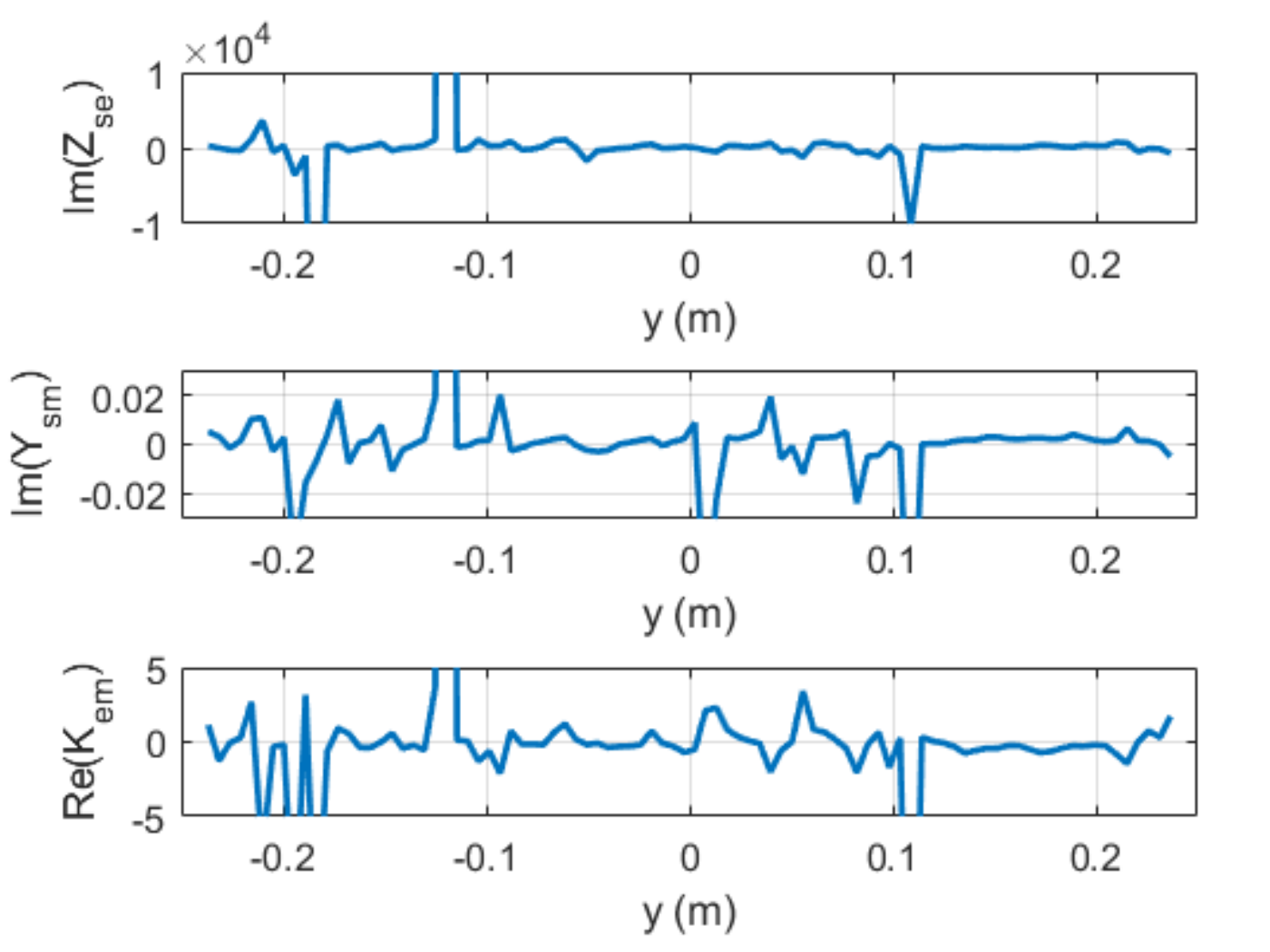}
  	\caption{(a) Desired ${Z}_{se}$, ${Y}_{sm}$, ${K}_{em}$ for transforming normal plane wave from one side of the EMMS to $45^\circ$-directed pencil beam. }
  	\label{fig:ZYK_refrac}
\end{figure}

\begin{figure}[!]
	\centering
  	\includegraphics[width=3.5in]{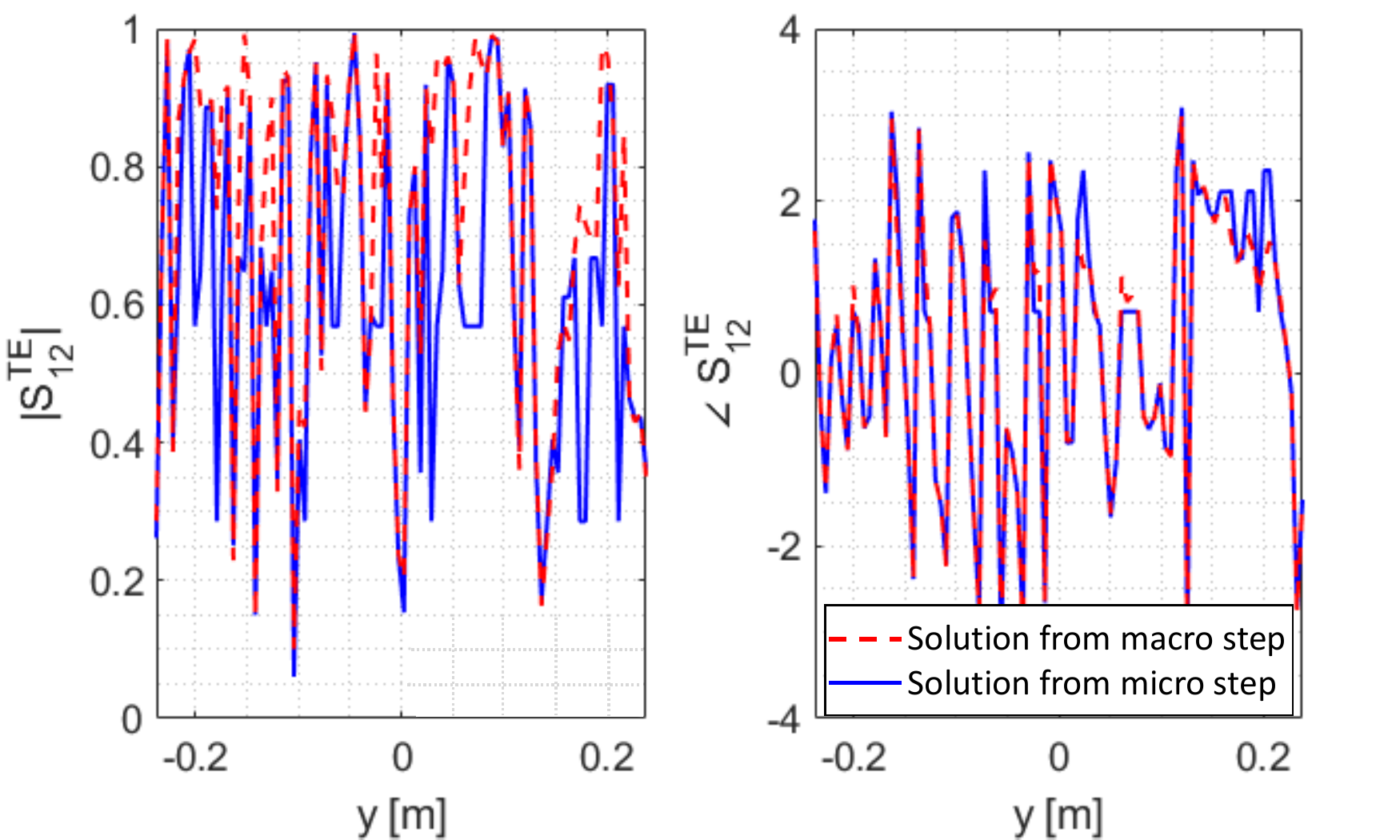}
  	\caption{$|S_{12}^{TE}|$ and $\measuredangle S_{12}^{TE}$ obtained from the macroscopic and microscopic optimizations for the first example. }
  	\label{fig:S_refrac}
\end{figure}

\begin{figure}[!h]
	\centering
  	\includegraphics[width=2.5in]{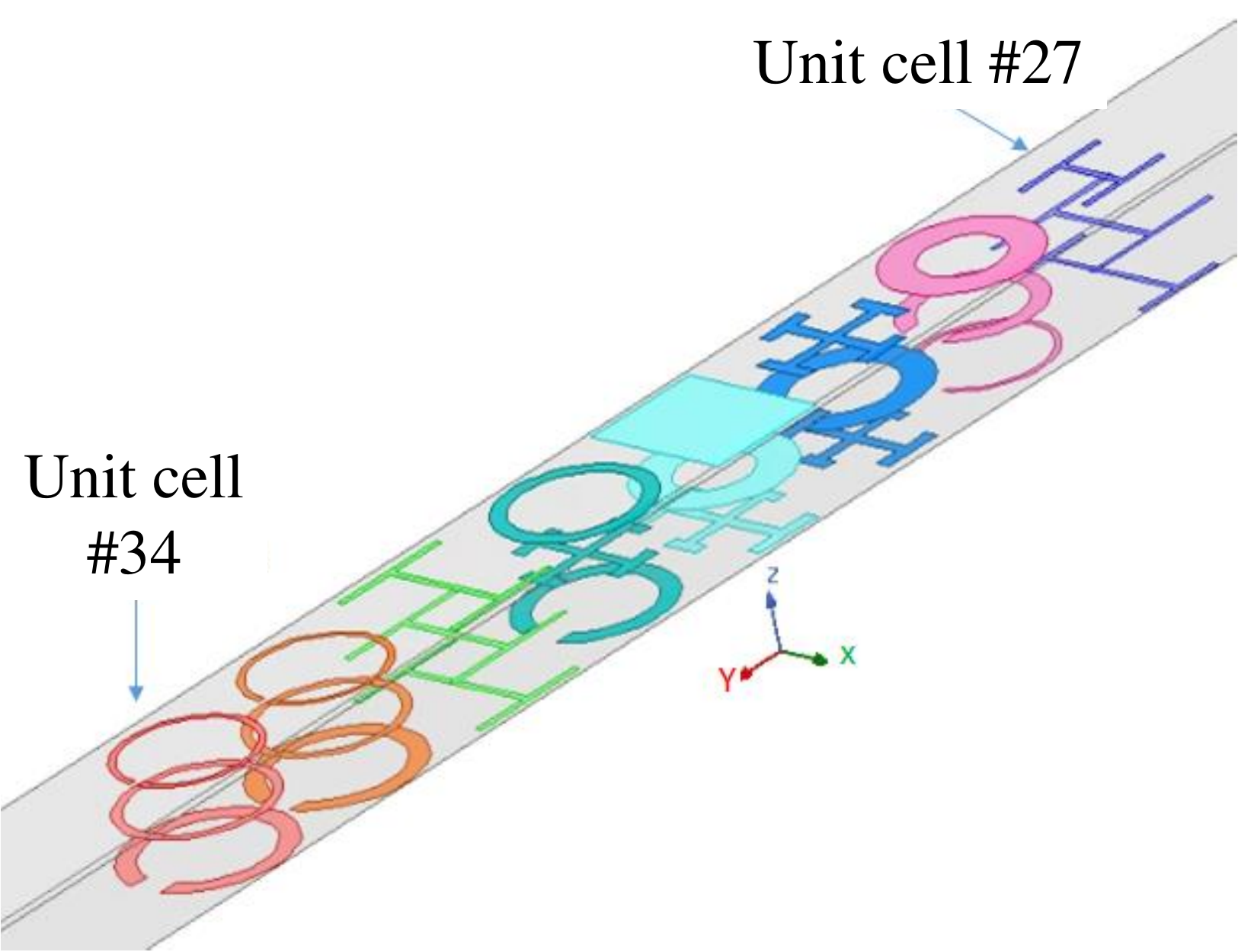}
  	\caption{A section of the optimized EMMS for example 1. }
  	\label{fig:cells_refrac}
\end{figure}

\begin{figure}[!]
	\centering
  	\includegraphics[width=3.6in]{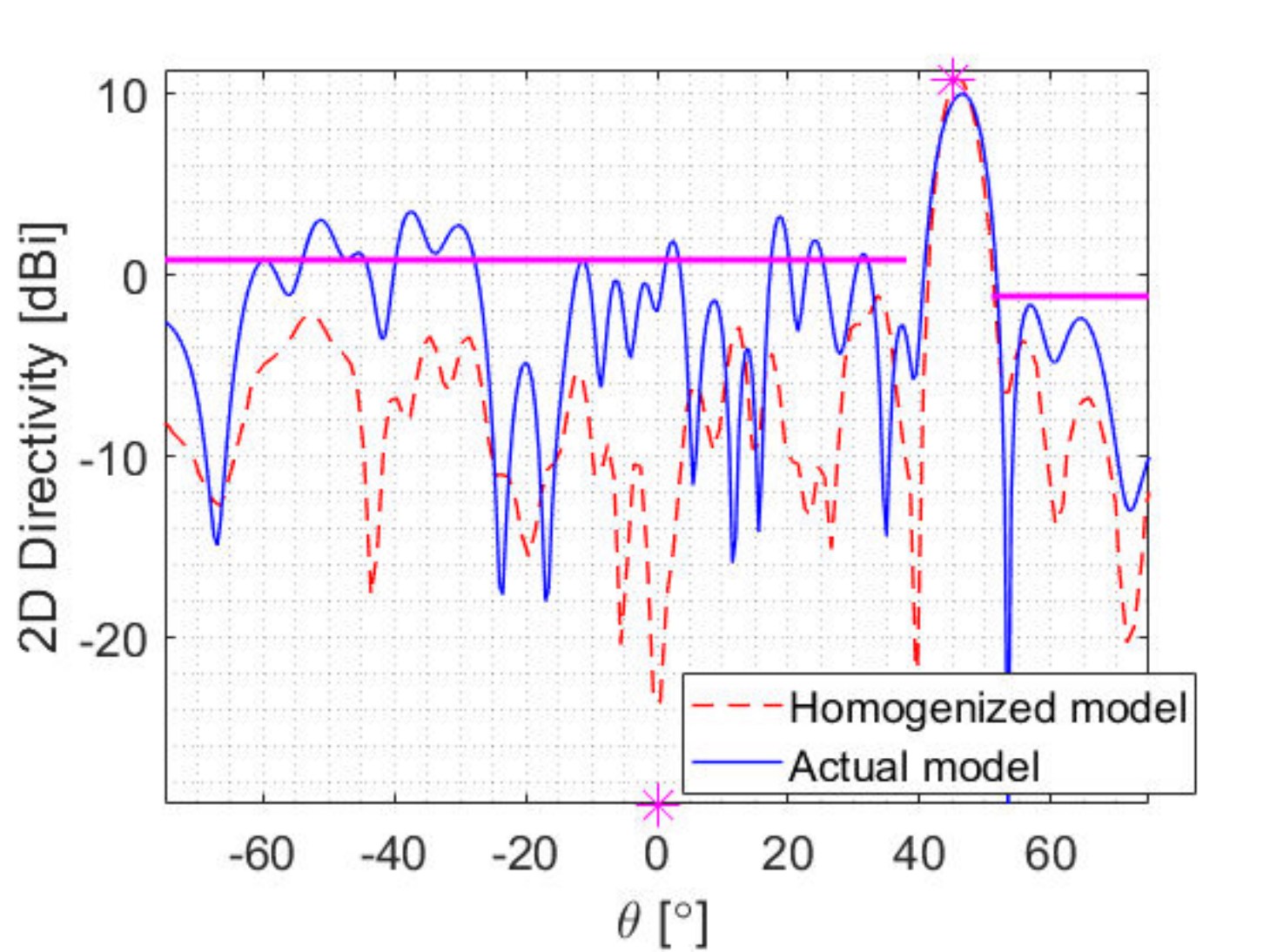}
  	\caption{Far-field (FF) patterns obtained from the homogenized model (dashed red line) and the actual model (blue solid line) simulated in Ansys HFSS for the specified main lobe, SLL, and null constraints (magneta) at $9.4$ GHz. }
  	\label{fig:FF_refrac}
\end{figure}

The desired $Z_{se} , Y_{sm},$ and $K_{em}$ for this transformation are shown in Figure \ref{fig:ZYK_refrac}. These macroscopic properties are converted to S-parameters to obtain $S_{11, desired}$ and $S_{12,desired}$. The PSO loss function based on these parameters is defined in \eqref{eq:LPSOmod}. To minimize this loss function, the sub-domain with $h^*=1.525$ mm is found as the sub-optimized domain to be explored as described in Section \ref{sec:III.D}. For this example as an initial test, we have simplified the problem and limited the solution space to the unit cells with non-complementary primitives on their middle layer but they are considered in the design space of the next example. Figure \ref{fig:S_refrac} shows the optimized $S_{12}$ from the macroscopic step and $S_{12}$ provided by the optimized physical unit cells in this domain. It can be seen that most of the desired properties are met with a small error. 

A subsection of the optimized EMMS is shown in Figure \ref{fig:cells_refrac}, where it is shown that the adjacent unit cells have completely different primitives. The optimized physical EMMS is simulated and the far-field radiation pattern is shown in Figure \ref{fig:FF_refrac}, where it is compared with the far-field obtained from the optimized $Z_{se} , Y_{sm},$ and $K_{em}$ based on the approach explained in Section \ref{sec:II}. The comparison between the optimized far-field and the imposed constraints shows that the macroscopic step is successful in outputting the optimized surface parameters. While the constraints are mostly met by the synthesized EMMS, the null region in the simulated far-field is not realized and the sidelobe levels are slightly higher than the desired level. This is due the fact that the mutual coupling between the unit cells has significant impact on both of these properties. This physical mutual coupling between the actual unit cell conductors is not fully captured by the inter-cell mutual coupling predicted by the homogenized model discussed in Section \ref{sec:II}. Despite this difference, the proposed approach is successful in matching physical unit cells to a wide variety of surface parameters and generating the desired radiation pattern.

\vspace{-0.3cm}

\subsection{Example 2: Multi-Beam Optimization} 

In the second example, we design an EMMS that transforms the fields from a line source located at $x_f = y_f = 0$ mm and $z_f=- D/4$ into an asymmetric multi-beam far-field radiation pattern. The details of the optimization problem and the far-field constraints are listed in Table \ref{MultiCrit_table} and \eqref{MultiCritOptProblem}.

\vspace{0.5cm}
\captionsetup{skip=2pt}
\begin{table}[!h]
\caption{Multi-Beam optimization parameters}
\label{MultiCrit_table}
\centering
\begin{tabular}{|c|c|}
\hline
 Parameter &  Value \\
\hline
 Incident  & $\frac{1}{\sqrt{r}}e^{-jkr}$ \\
 Field  ($E^{inc}$)  & $r = \sqrt{y^2+(D/4)^2 }$ \\ 
\hline
 Angular Sampling Points ($M$) & $361$ \\
\hline
 Spatial Sampling Points ($N$) & $90$ \\
\hline
 Max Iterations & $295$ \\
\hline
 Initial & $([{\textbf{X}}_{se}]^0,[{\textbf{B}}_{sm}]^0,[{\textbf{K}}_{em}]^0) = \textbf{0}, $\\
 Conditions & $\rho = 10,(\bm{\mu}_{\textbf{Z}_e}^0, \bm{\mu}_{\textbf{Z}_m}^0) = \textbf{0}$ \\
\hline
 Main Lobe Angles  ($MB$) & $\theta = \left\lbrace 30^{\circ},-20^{\circ}\right\rbrace$ \\
\hline
 Main Lobe  & $\lbrace 10.65+j10.65, \theta = 30^{\circ} \rbrace$, \\
 Level ($MB_{level}$) & $\lbrace 12.56+j12.56, \theta = -20^{\circ} \rbrace$  \\
\hline
 Sidelobe  & $ \lbrace  3.85,0^{\circ}\leq \theta\leq 11^{\circ}\rbrace ,$\\
 Level ($\bm{\tau}$)& $ \lbrace  3.95,  49^{\circ}\leq \theta\leq 90^{\circ}\rbrace ,$ \\
 and Angles ($SL$)  &  $ \lbrace 2.55,  90^{\circ}\leq \theta\leq 180^{\circ}\rbrace , $\\
 & $  \lbrace 2.55,  -180^{\circ}\leq \theta\leq -90^{\circ}\rbrace ,$\\
 &  $ \lbrace 3.85,  -90^{\circ} \leq \theta\leq -36^{\circ}\rbrace ,$\\
 & $ \lbrace 3.95,  -4^{\circ} \leq\theta\leq 0^{\circ}\rbrace$\\
\hline
 Null Angles ($NU$)& $\theta=\left\lbrace 0^{\circ},180^{\circ}\right\rbrace$  \\
\hline
\end{tabular}
\end{table}

\begin{subequations} \label{MultiCritOptProblem}
\begin{align}
\minimize_{\substack{\textit{\textbf{I}}^e,\textit{\textbf{I}}^m,[\textbf{X}_{se}],[\textbf{B}_{sm}],[\textbf{K}_{em}],\\slack_{D^e},slack_{D^m}}} \begin{split}&\quad 10000f_{MB}(MB)+\\&\quad 100f_{NU}(NU)+ f_{D}\end{split}\\
\textrm{subject to}  \quad   \begin{split}& \quad {\textit{\textbf{E}}}^{inc} = [{\textbf{Z}}_e]{\textit{\textbf{I}}}^e+[{\textbf{X}}_{se}]{\textit{\textbf{I}}}^e\\&\quad -[{\textbf{K}}_{em}]{\textit{\textbf{I}}}^m \end{split}\\
\begin{split}& \quad 1000({\textit{\textbf{H}}}^{inc} = [{\textbf{Z}}_m]{\textit{\textbf{I}}}^m\\ & \quad +[{\textbf{B}}_{sm}]{\textit{\textbf{I}}}^m +[{\textbf{K}}_{em}]{\textit{\textbf{I}}}^e)\end{split} \label{H_eqn}\\
\begin{split}&\quad\vert [{\textbf{G}}^e](SL){\textbf{\textit{I}}}^e+[{\textbf{G}}^m](SL){\textbf{\textit{I}}}^m \\ &\quad+ {\textbf{E}}_{ff}^{inc}(SL) \vert \leq \bm{\tau} \end{split} \\
& \quad \vert [\textbf{D}]{\textbf{\textit{I}}}^e\vert \leq \textbf{0.1} + slack_{D^e} \\
& \quad \vert [\textbf{D}]{\textbf{\textit{I}}}^m\vert \leq \textbf{25}+ slack_{D^m},
\end{align}
\end{subequations}

\begin{figure}[!h]
	\centering
  	\includegraphics[width=3.5in]{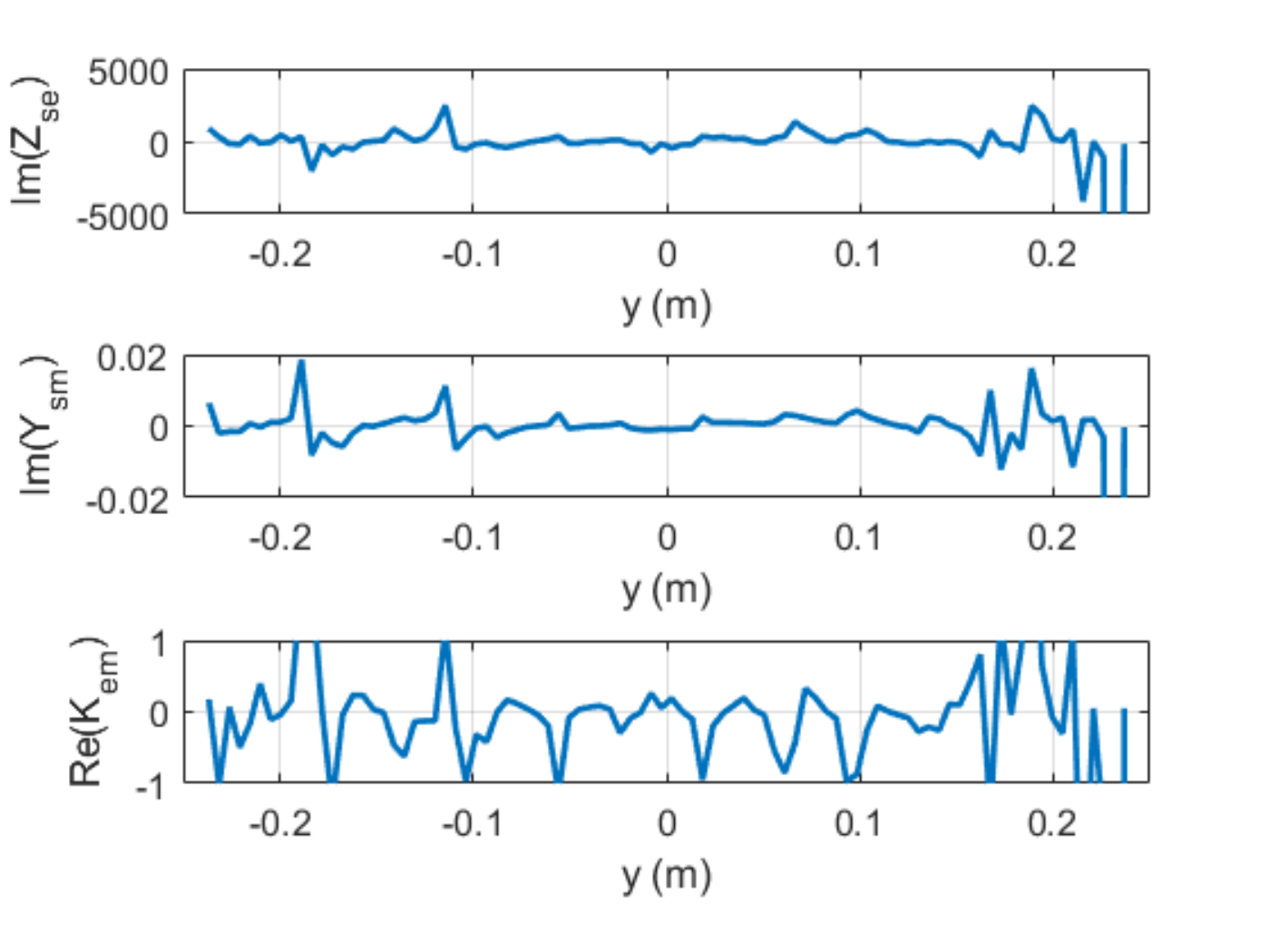}
  	\caption{ Desired ${Z}_{se}$, ${Y}_{sm}$, ${K}_{em}$ for transforming cylindrical wave from the line source located at $F=D/4$ from one side of the EMMS to the multi-beam far-field on the other side.}
  	\label{fig:ZYK_multi}
\end{figure}

\begin{figure}[!]
	\centering
  	\includegraphics[width=3.5in]{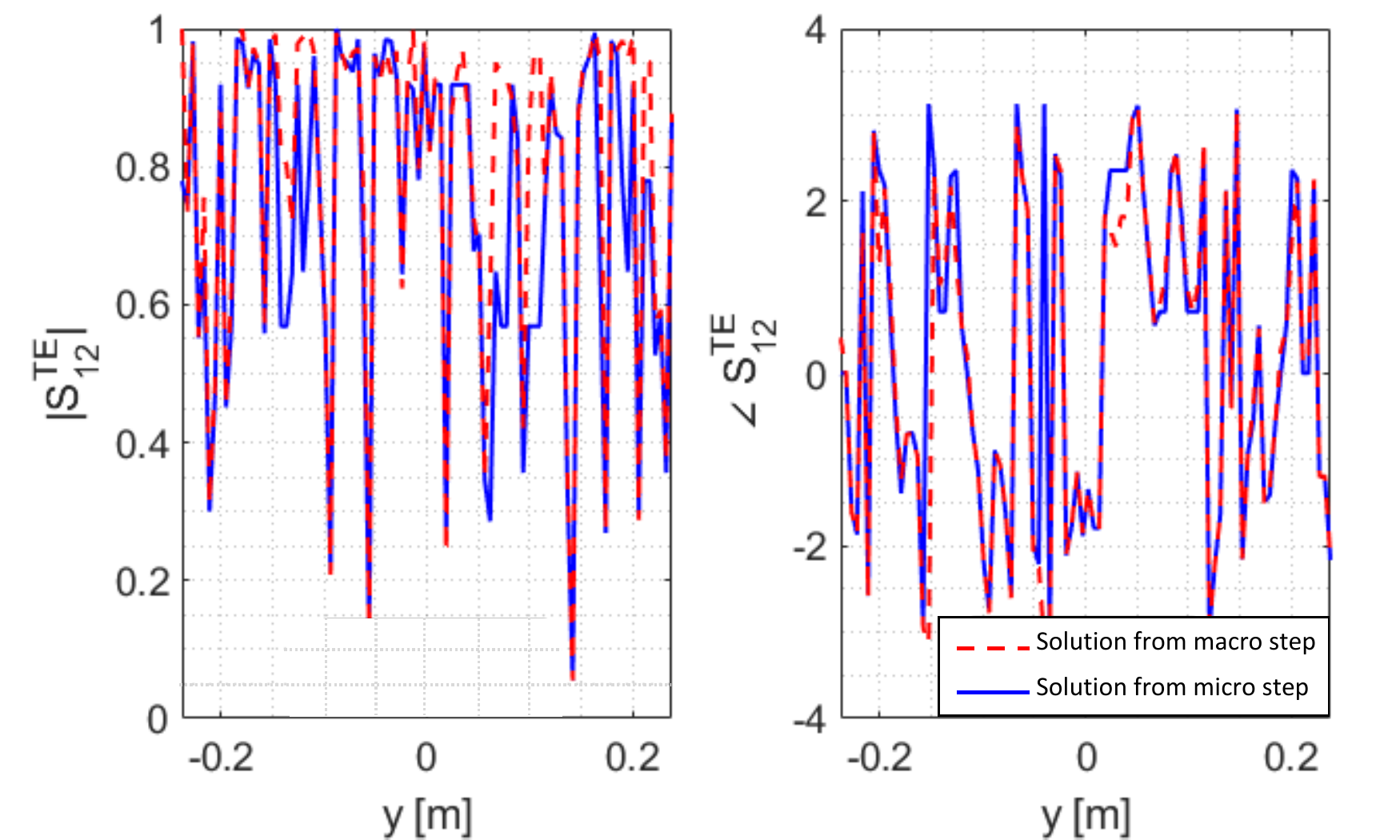}
  	\caption{$|S_{12}^{TE}|$ and $\measuredangle S_{12}^{TE}$ obtained from the macroscopic and microscopic optimizations for the second example. }
  	\label{fig:S_multi}
\end{figure}

\begin{figure}[!]
	\centering
  	\includegraphics[width=2.5in]{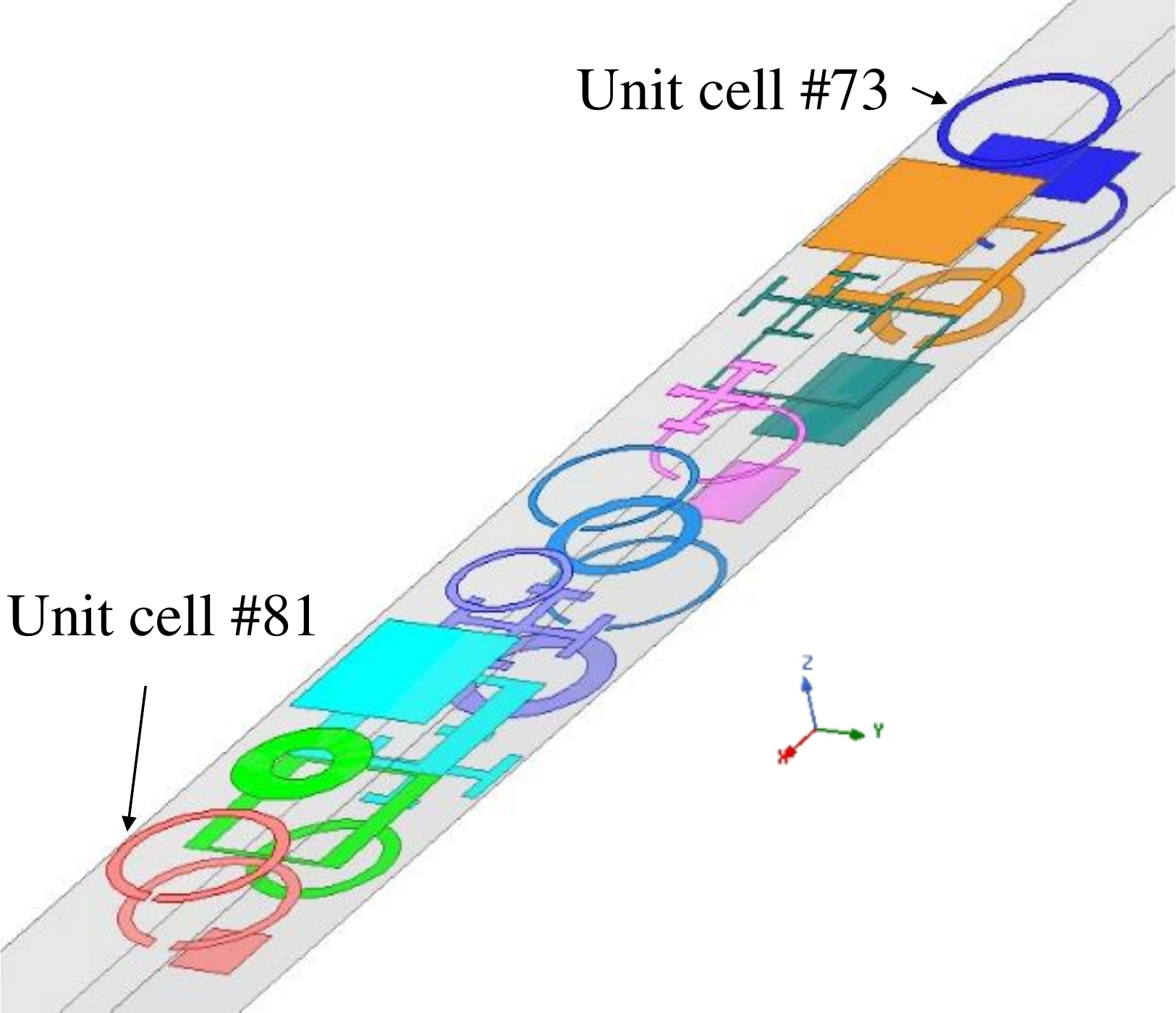}
  	\caption{A section of the optimized EMMS for example 2. }
  	\label{fig:cells_multi}
\end{figure}

Figure \ref{fig:ZYK_multi} shows the ${Z}_{se}$, ${Y}_{sm}$, and ${K}_{em}$ that are the optimized solutions to the problem described in \eqref{MultiCritOptProblem}. Figure \ref{fig:S_multi} shows the desired $|S_{12}^{TE}|$ and $\measuredangle (S_{12}^{TE})$ that are converted from the desired ${Z}_{se}$, ${Y}_{sm}$, and ${K}_{em}$. Based on these desired scattering parameters, the domain with the substrate thickness $h^*=1.525$ mm was chosen as the sub-solution space to be explored. Unlike the previous example, in this example to meet the desired surface properties, the domains with both complementary and non-complementary primitives on the middle layer are considered. The $|S_{12}^{TE}|$ and $\measuredangle (S_{12}^{TE})$ provided by the optimized physical unit cells are shown in Figure \ref{fig:S_multi}, where they are in excellent agreement with the optimized parameters from the macroscopic step. Similar to the previous example, the optimized EMMS here is also composed of different primitives of scatterers. A subsection of the array is shown in Figure \ref{fig:cells_multi}, where it can be seen that the unit cells $74-75$ and $79-80$ have complementary rectangular patch (compRP) scatterers on their middle layers.

\begin{figure}[!]
	\centering
  	\includegraphics[width=3.6in]{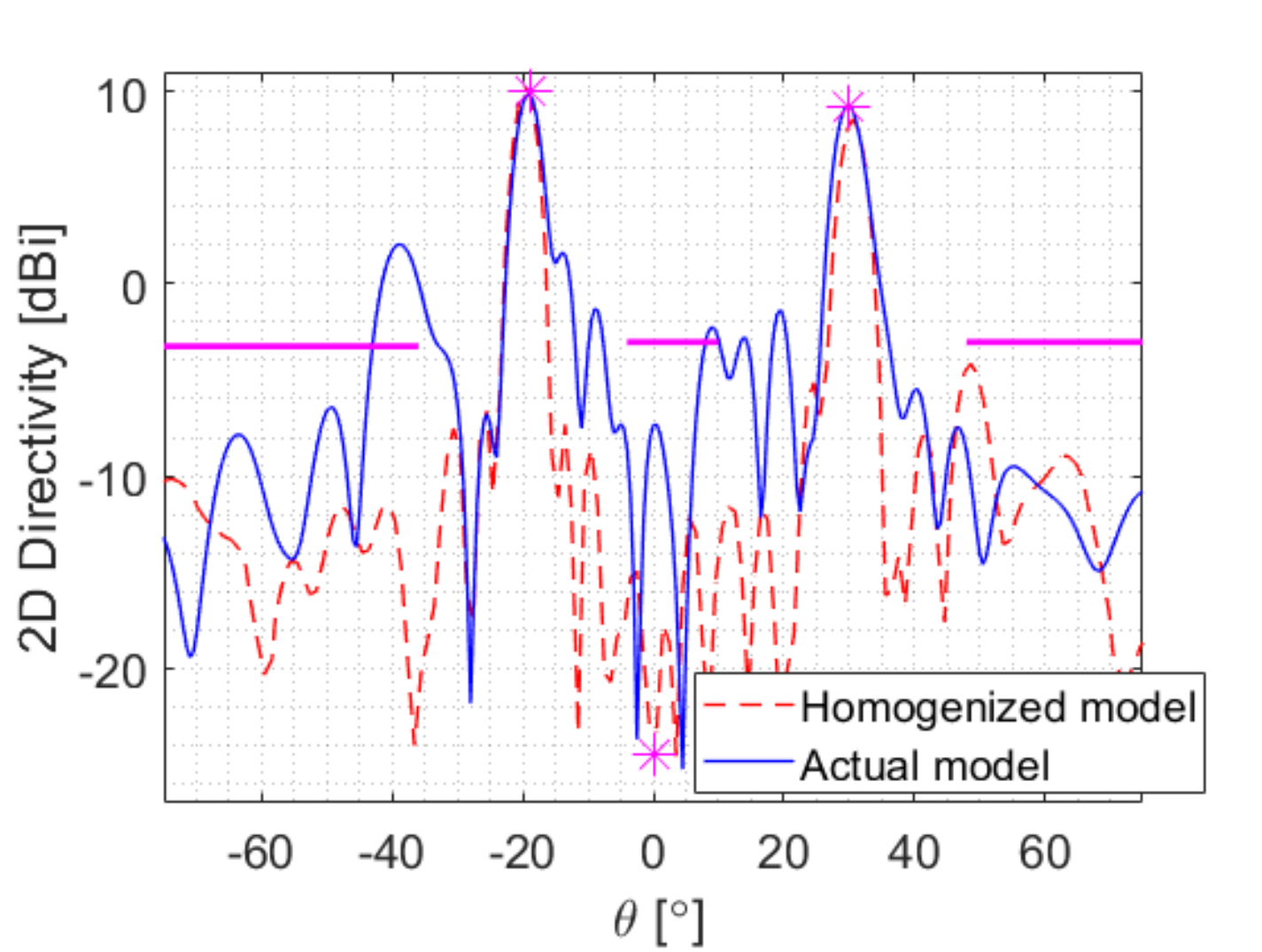}
  	\caption{Far-field (FF) patterns obtained from the homogenized model (dashed red line) and the actual model (blue solid line) simulated in Ansys HFSS for the specified main lobe, SLL, and null constraints (magneta) at $9.4$ GHz. }
  	\label{fig:FF_multi}
\end{figure}

Figure \ref{fig:FF_multi} shows the far-field pattern calculated from the desired ${Z}_{se}$, ${Y}_{sm}$, and ${K}_{em}$ shown in Figure \ref{fig:ZYK_multi}, the applied constraints, and the simulated far-field pattern from the physical optimized EMMS. Despite the fact that the optimized constituent unit cells have different primitives and structures, it can be seen that the optimized EMMS produces a far-field that satisfies  the constraints everywhere except the maximum sidelobe level for $\theta \in [-43, -35] ^ \circ$. It is worth mentioning that this is due to the fact that the behavior of the unit cells in the non-periodic optimized EMMS deviates from their properties with the periodic boundary condition, due to the additional mutual coupling between cells that are not modeled by the inter-cell coupling in the homogenized model. Nonetheless, through this example, we have shown that the proposed approach can be successfully employed for the inverse design of a nonuniform bianistropic metasurface that transforms the excitation from a line source to a complex far-field pattern comprising two main beams with different radiation intensities.

\section{Conclusion} \label{sec:Conclusion}
In this paper, the inverse design problem of a multilayer nonuniform metasurface based on the high-level far-field constraints is solved in a systematic and end-to-end approach for the first time. We divide this problem into two  inverse problems: a macroscopic problem and a microscopic problem, where a combination of machine learning and optimization techniques is employed to first obtain the optimized surface parameters and then the physical EMMS. This problem has many degrees of freedom including  scatterer primitives and dimensions and can have more than one solution. Therefore, instead of using \textit{ad hoc} and heuristic methods, we implement the alternating direction method of multipliers (ADMM)-based convex optimization to deterministically obtain the EMMS surface parameters derived from the method of moments (MoM) to satisfy constraints on the main lobe level(s), sidelobe levels, and null positions on the radiated far-field. Then, the particle swarm optimization integrated with deep neural networks used as surrogate models is employed to explore and exploit the physical solution space for the optimized EMMS. 

Compared to the image-based representations of unit cells \cite{b20}, the proposed method to represent 3-layer bianisotropic unit cells is more successful in capturing both the categorical and continuous nature of the scatterers. Hence, it can interpolate the scattering properties of individual layers and interlayer coupling in a new unit cell based on the samples in the training data with reduced error, which removes the need for further full-wave simulations. The surrogate models expedite the exploration through the solution space so much that the designer can choose large swarms or number of iterations for the PSO and find the global optimum. 

Two passive and lossless nonuniform bianisotropic metasurfaces are designed to transform the excited fields from line source(s) to complex far-fields that successfully satisfy the constraints on main lobe(s), sidelobes, and null regions. The constituent unit cells of the optimized EMMSs can have completely different primitives from their neighbors, which is necessary to match the desired optimized surface parameters but it also results in discrepancies of the results due to the unpredictable levels of additional mutual coupling introduced beyond that captured in the homogenized model. In the examples we have shown, the mutual coupling has mostly impacted the null realization and sidelobe levels of the far-field radiated from the physical EMMS with acceptable error. Nevertheless,  the presented results demonstrate the effectiveness of the proposed approach in the inverse design of different EMMSs for various types of application.   

Despite the demonstrated success of the method, several practical considerations must be mentioned. First and foremost, the macroscopic optimizer described in Section \ref{sec:II} must be supplied with a roughly feasible problem to optimize for if the designer is to meet their objectives. There are physical limitations to the directivity, sidelobe level, number of beams, etc. that an EMMS can realize. Although the macroscopic optimizer can be used as a rough guide for the feasibility of a problem (i.e. if it converges or not) it is not rigorous. The macroscopic optimizer might also return optimized surface parameters that may not be achievable from a 3-layer unit cell given the constraints on substrate thicknesses and permittivities. Designers might also need to include scatterers with more complex primitives and a wider range of scattering properties in the solution space to match the desired surface properties with a better accuracy. Furthermore, if the surface parameters change abruptly and frequently across the surface, it means that the physical structure of the EMMS also needs to follow the properties, thereby causing more uncertainty due to the mutual coupling. To circumvent this issue, limiting constraints on the spatial derivatives of the surface parameters can be added to the  macroscopic optimization problem, which will be considered in the future.

There remain some areas for future research to augment the utility of this method. Firstly, a method to automatically determine the optimization weights of the macroscopic optimizer would reduce the amount of tuning required. Secondly, a minimum level field constraint would also be very useful. This would enable the specification of minimum directivity levels for beams or more sophisticated patterns (such as isoflux patterns). Thirdly, moving this design scheme into three dimensions would provide more degrees of freedom to optimize for objectives at different elevations. Finally, a fully generative machine-learning model \cite{b20} combined with generalized reliable surrogate models can be integrated into the microscopic optimization step to efficiently explore the physical solution space of 3-layer bianisotropic unit cells to a greater degree and achieve desired surface parameters with reduced error.


%

\ifCLASSOPTIONcaptionsoff
  \newpage
\fi



%

\end{document}